\documentclass{aa}
\usepackage{graphics}
\begin{document}
\thesaurus{}
\title {Spectroscopy  of the post-AGB star 
HD~101584(IRAS~11385-5517)
\thanks{Based on observations obtained 
 at the European Southern Observatory(ESO), 
Chile and  the  Vainu Bappu Observatory,  Kavalur, India}}
\author{ T.~Sivarani \inst{1} 
\and M.~Parthasarathy \inst{1} 
\and P.~Garc\'\i a-Lario \inst{2} 
\and A.~Manchado \inst{3} 
\and S.R.~Pottasch \inst{4}}
\institute{Indian Institute of Astrophysics, Bangalore 560 034, India. 
\and ISO Data Centre, Astrophysics Division, Space Science Department of
ESA, Villafranca del Castillo, Apartado de Correos 50727,28080 Madrid, Spain.
\and Instituto de Astrofisica de Canarias, E-38200 La Laguna, Tenerife, Spain.
\and Kapteyn Astronomical Institute, Postbus 800, NL-9700 AV Groningen, The Netherlands. }
\offprints{T. Sivarani}
\mail{sivarani@iiap.ernet.in}
\date{Received /Accepted }
\maketitle
\titlerunning{HD~101584}
\authorrunning{T. Sivarani et~al.}
\markboth{HD~101584}{}

\begin{abstract}

From an analysis of the spectrum (4000\AA\ to 8800\AA) of HD~101584
it is found that most of the neutral and single ionized metallic
lines are in emission. The forbidden emission lines  of [OI] 
6300\AA\ and 6363\AA\ and [CI] 8727\AA\ are detected, which 
indicate the presence of a very low excitation nebula. 
The H$\alpha$, FeII 6383\AA, NaI D$_{1}$, D$_{2}$ lines and the 
CaII IR triplet lines show P-Cygni profiles indicating a mass 
outflow. The H$\alpha$ line shows many velocity components in 
the profile. The FeII 6383\AA\ also has almost the same line profile
as the H$\alpha$ line indicating that they are formed in the  
same region. From the  spectrum synthesis analysis we find the
atmospheric parameters to be T$_{eff}$=8500K, log g=1.5,
V$_{turb}$=13km~s$^{-1}$ and [Fe/H]=0.0.
From an analysis of the absorption lines the photospheric 
abundances of some of the elements are derived. 
Carbon and nitrogen are found to be overabundant.
From the analysis of Fe emission lines  we 
derived T$_{exi}$~=~6100K$\pm$200  for the emission line region.

\end{abstract}

\section{Introduction}
Humphreys and Ney (1974) found near-infrared excess in HD~101584 and 
suggested that it  is a massive F-supergiant with an M-type
binary companion star (Humphreys 1976). However, HD~101584 (V=7.01, F0 Iape
(Hoffleit et al. 1983)) 
was found to be an IRAS source (IRAS~11385$-$5517) (Parthasarathy and
Pottasch 1986). On the basis of its far-infrared colors, flux
distribution and detached cold circumstellar dust shell, Parthasarathy 
and Pottasch (1986) suggested that  it is a low mass star in the
post-Asymptotic Giant Branch (post-AGB) stage of evolution. 

       CO molecular emission lines at millimeter wavelengths were detected 
by Trams et al. (1990). 
The complex structure of the CO emission shows large Doppler velocities 
of 130 km~s$^{-1}$ with respect to the central velocity of the feature
indicating a very high outflow velocity. 
Te Lintel Hekkert et~al. (1992) reported  the discovery of OH 1667 MHz 
maser emission from the circumstellar envelope of HD~101584. The OH 
spectrum has a velocity range of 84 km~s$^{-1}$ and shows two 
unusually broad emission features. Te Lintel Hekkert et~al. (1992) 
found from the images obtained from the Australian  Telescope, that the
OH masers are located along the bipolar outflow.
 The post-AGB nature of HD~101584
is also suggested by the space velocity of the star derived from 
the central velocity of the CO and OH  line emission.
This velocity of V$_{rad}$~=~50.3 $\pm$ 2.0 km~s$^{-1}$ does not agree 
with the galactic rotation curve assuming it to be a luminous massive 
population I F supergiant.

      Bakker et~al. (1996a) studied the low and high resolution ultraviolet spectra 
and  the high resolution optical spectra of HD~101584. Based on the strength of HeI
(see also Morrison and Zimba 1989) N~II, C~II lines and Geneva photometry,
Bakker et~al. (1996a) suggest  that HD~101584 is a B9~II star of 
T$_{eff}$~=~12000K $\pm$ 1000K and log g~=~3.0. Bakker et~al. (1996b)
also found small amplitude light and velocity variations and suggested that 
HD~101584 is a binary with an orbital period of 218 days.

      The optical spectrum of HD~101584 is very complex and shows  many lines 
in emission. In this paper we report an analysis of the high resolution 
optical spectrum of HD~101584.

\section {Observations and analysis}
High resolution and high signal to noise ratio spectra of HD~101584 were 
obtained with the European Southern Observatory (ESO) Coude Auxiliary
Telescope (CAT) equipped with the Coude Echelle Spectrograph (CES) and 
a CCD as detector. The spectra cover the wavelength regions 5360-5400\AA, 
6135-6185\AA, 6280-6320\AA, 6340-6385\AA, 6540-6590\AA, 7090-7140\AA, 
7420-7480\AA, 8305-8365\AA\ and 8680-8740\AA. The spectral resolution 
ranged from 0.165\AA\ at 6150\AA\ to 0.210\AA\ at 8700\AA.
 We have also 
obtained 2.5\AA\ resolution spectra of HD~101584 from 3900\AA\ to 8600\AA\ 
with the 1m telescope and UAGS spectrograph and a  CCD as detector at the 
Vainu Bappu Observatory (VBO), Kavalur, India. In addition we obtained 
CCD spectra with the same telescope and Coude Echelle spectrograph,
covering the the wavelength region 4600\AA\ to 6600\AA\ with a resolution
of 0.4\AA. All spectra mentioned above were used in this analysis.

 All the spectra were analyzed using IRAF software.
The equivalent widths of lines were found by fitting a
gaussian. For blended lines de-blending was done by fitting multiple gaussians.
We carried out spectrum synthesis calculations using KURUCZ  stellar models (1994).
SYNSPEC code (Hubeny et al. 1985) was used for calculating the theoretical line profiles. 
The gf values were taken from Wiese et~al. (1966), Wiese and Martin (1980), Hibbert et~al.(1991), 
Parthasarathy et al. (1992) and Reddy et al.  (1997 and references therein).
For the analysis of forbidden lines we have used the IRAF software package
NEBULAR under STSDAS.

\section {Description of the spectrum}
 The remarkable characteristic of the optical spectrum of HD~101584 is 
the fact that different spectral regions  resemble different spectral
types. The spectrum in the UV region is similar to that of $\alpha$ Lep
which is an F-supergiant (Bakker 1994). The optical spectrum in the range
3600\AA-5400\AA\ is dominated by absorption lines. Most of them are due to
neutral and single ionized lines of Ti, Cr and Fe.  
The CaII H and K absorption lines are strong. 
The strength of the absorption lines are similar to that observed 
in an A2 supergiant. In the yellow and red spectral regions, most of 
the lines are in emission (Fig. 1). 
The emission lines show complex line profiles.
The absorption lines of NI, OI, CII and SiII are broad. The
Paschen lines are in absorption. Some of these absorption lines
are blended with emission lines and many have asymmetric profiles. 
The OI lines at 6156\AA\ are blended with emission lines of FeI. 
The NI lines are strong and show asymmetric line profiles. 
The blue wing is shallow compared to the red wing. The CII 
lines at 6578\AA\ and 6582\AA\ are weak. The Na D lines, 
KI 7700 \AA\ (Fig. 2), the CaII IR triplet lines (Fig. 3), [OI], [CI] and 
MgI 6318.7\AA\ lines are found in emission. The OI triplet lines (Fig. 2)
are very strong indicating an extended atmosphere and NLTE effects.

\subsection{P-Cygni profiles}
The H$\alpha$ line has a very strong P-Cygni profile indicating an outflow. 
The profile looks very complex. It shows at least 6 velocity components.  
The FeII line at 6383\AA\ is in emission and the profile is very
similar to  that of H$\alpha$ (Fig. 4). Similar behaviour of the 6383\AA\
FeII line and H$\alpha$ line is also noticed in the post-AGB F supergiant
IRAS 10215-5916 (Garc\'\i a-Lario et al. 1994). The H$\alpha$ and the FeII 6383\AA\
line show an outflow velocity of 100$\pm$10 km~s$^{-1}$.
The H$\beta$ line also shows a P-Cygni profile. It has a broad 
emission wing at the red end. This indicates that the line forming
region is extended.  The H$\beta$, NaI D1, D2 and the CaII IR triplet lines (Fig. 3) 
show an outflow velocity of 75$\pm$20 km~s$^{-1}$.
 The velocity structure seen in these P-Cygni profiles
could be due to emission from different shells formed during the episodic
mass-loss events.

\subsection{FeI and FeII emission lines}
 The presence of numerous emission lines of FeI and FeII makes it
possible to derive the physical conditions of the line forming region.
From the curve of growth analysis of the FeI and FeII emission lines (Viotti 1969),
we have derived T$_{exi}$=6300$\pm$1000K  and 5550$\pm$1700K respectively (Fig. 5). 
The scatter found could
be due to the fact that the lines are not optically thin. On the other hand,
there are only few emission lines of FeII present in the spectra and thus
the estimate from FeII might not be accurate.
In order to determine whether the large scatter observed in Fig. 4 is reflecting
optical thickness effects we have done self-absorption curve (SAC) analysis 
(Friedjung and Muratorio 1987) for the FeI emission lines. 

   SAC is a kind of curve of growth applied to emission lines, but it has certain
advantages as compared to the classical emission line curve of growth analysis.
This method of analysis is valid also for optically thick lines. It deals with
each transition separately, so that it is possible to get the population 
of different levels without assuming a Boltzmann distribution.
In this curve, a function of the line flux emitted in the different transition
of a given multiplet is taken in such a way that it is constant for a optically
thin uniform medium. As the optical thickness increases the curve will move
towards a straight line inclined at -45$^{o}$.
   The shape of the SAC in Fig. 6(a) shows the lines are optically thick. 
The shape of the SAC is obtained by shifting all the multiplets with respect
to a reference multiplet. Here we have taken multiplet 207 as reference.
The X and Y shifts of each multiplet gives the relative population of the 
lower  and upper level with respect to the reference multiplet. 
Fig. 6(b) and Fig.6(c) shows the Y and X shifts versus the 
upper and lower excitation potential from which we derive the 
T$_{exi}$=6100$\pm$200K.

\subsection{Forbidden lines}
The forbidden emission lines at 5577\AA, 6300\AA\ and 6363\AA\ of neutral oxygen
are present in the spectra. The forbidden line of neutral carbon at 8727\AA\
is also seen. The 6300\AA\ line is blended with ScII line and the 5577\AA\ 
line is very weak. We have calculated the  I(6300)+I(6363)/I(5577)
to be 13.3. From the flux ratio we can calculate T$_{e}$ (Osterbrock 1989). 
This flux ratio is not very accurate because of very the weak 5577\AA\
line and poor signal to noise spectrum.
For the flux ratio of 13.3 we derived a function 
depending on the electron density N$_{e}$ and temperature T$_{e}$. 
Figure 7 shows the N$_{e}$ and T$_{e}$ contours for different values of
flux ratio around 13.3. Since we do not
see any other forbidden lines which are sensitive to the electron density,
we could not fix both N$_{e}$ and T$_{e}$ uniquely.  But assuming a temperature
derived from the Fe emission lines, an electron density of  1 x 10$^{7}$ is obtained.
For this value of electron density and temperature the C/O = 0.5$\pm$ 0.2 
has been obtained.

\section{Radial velocities}
There are very few absorption lines and most of these are affected by 
emission and or a shell component, therefore we derived the average
radial velocity from the well defined emission lines.
The average radial velocity from the emission lines is found to 
be 50 $\pm$ 2km~s$^{-1}$.
Morrison and Zimba (1989) using 14 best absorption lines found the 
radial velocity to be 69$\pm$ 1 km~s$^{-1}$.
From the equivalent widths of FeI absorption lines given by 
Rosenzweig et~al. (1997) we find  no correlation 
between log gf - $\chi$$\Theta$ and heliocentric radial 
velocity (Fig. 8). 
However, Bakker et~al. (1996a) found a correlations between 
log gf - $\chi$$\Theta$ 
and heliocentric radial velocities of HD101584 in the UV. 
The discripancy could be due to the poor resolution of Rosenzweig et~al. (1997)
data compared to that of Bakker et~al. (1996a). 
The large scatter seen in the radial velocities could  be 
due to pulsation. 
Similar velocity variations were noticed in other post-AGB
supergiants (Garc\'\i a-Lario et al. 1997, Hrivnak 1997).

\section{Atmospheric parameters and chemical composition}

           The UV (IUE)  low resolution spectrum of HD~101584 
matches well with that of an A6Ia star (HD~97534) (Fig. 10) 
indicating a T$_{eff}$ of 8400K (Lang  1992). The presence of CII lines 
at 6578\AA\ and 6582\AA\ indicates
a T$_{eff}$ $>$ 8000K. 
For T$_{eff}$ $\le$ 8000K the CII lines would be very weak or absent.
The Paschen lines also indicates a
low gravity (Fig. 11). The luminosity class Ia also indicates a
very low gravity. From the analysis of several nitrogen lines
around 7440\AA\ and 8710\AA\ we derived the microturbulence
velocity V$_{turb}$=13km~s$^{-1}$. We
synthesised the spectral region from 4000\AA\ to 4700\AA\ (Fig. 9)
with low gravity (log  g =1.5 )models of Kurucz (1993) 
with temperatures 8000K, 8500K and 9000K. The best fit was found
for T$_{eff}$ = 8500K, log g = 1.5, V$_{T}$=13km~s$^{-1}$
and [Fe/H] = 0.0.
        
      The line at 5876\AA\ was identified as a HeI line by Bakker et al.(1996a)
who also state that the lines at 5047\AA\ and 5045\AA\ as due to HeI and NII
respectively. However, we find that the 5047 and 5045 lines are in fact due to FeII.
Except HeI 5876\AA, we have not found any other helium lines in the spectrum
and nor have we  found any NII or OII lines. In fact, Hibbert et al. (1991)
indicate the presence of a CI line at 5876\AA. It is likely that the line
at 5876\AA\ may be due to CI instead of HeI.

If we assume that the 5876\AA\ line is due to HeI then for a 
solar helium abundance and log g = 1.5, T$_{eff}$ = 9000K is found. 
Since we do not see any other helium lines, if 5876\AA\ line
is due  to helium , it is likely that it may be formed in the
stellar wind or in the chromosphere of the star.
On the basis of the presence of this  helium line 
Bakker et al. (1996a) suggested 
that HD~101584 is a B9II star of T$_{eff}$ 12000K. On the basis
of the analysis of our spectra we have not found any evidence for
such a high  temperature. 
We have also analysed the equivalent widths of absorption 
lines in the spectrum of HD~101584 given by Bakker et al.(1996a).
The final abundances
of some of the elements are listed in Table 2. The abundances
listed in Table 2 show that the star is overabundant
in carbon and nitrogen. It appears that the material processed
 by the triple alpha 
C-N and O-N cycle has reached the surface.

\section{ Discussion and Conclusions}

    The optical spectrum of the post-AGB star HD~101584 is rather
complex. We find several emission lines and P-Cygni profiles
indicating an ongoing mass-loss and the presence of a
circumstellar gaseous envelope. From the analysis of the 
absorption lines we find the atmospheric parameters to be
T$_{eff}$=8500K , log g=1.5, V$_{t}$=13km~s$^{-1}$
and [Fe/H]=0.0. 

        Carbon and Nitrogen are found to be overabundant
indicating that material processed by triple alpha C-N
and O-N cycles has reached the surface. Since our blue
spectra are of relatively low resolution and because of 
the presence of emission and shell components it is difficult to 
estimate reliable abundances of s-process elements.
The OI line at 6156\AA\ is blended with a weak FeI emission line.
The OI triplet at 7777\AA\ is very strong and affected by NLTE.
In any case it appears that the oxygen abundance is nearly solar.
A NLTE analysis of the high resolution OI 7777\AA\ triplet
may yield a more reliable oxygen abundance. 
 
      The nitrogen abundance is based on 6 lines in the 7440\AA\
and 8710\AA\ region. We have not used the strong nitrogen lines.
Nitrogen seems to be clearly overabundant. The carbon abundance 
is based on two CII lines at 6578\AA\ and 6582\AA. There is
a clear indication that carbon is overabundant. 
The abundance of Mg, Ti, and Fe  are nearly solar. The Ti abundance
is based on 15 lines and the Fe abundance is based on 6 lines.
Many of the other atomic lines are affected by emission 
and shell components. 
     In our opinion, the line at 5876\AA\ might be due to CI
(Hibbert et~al. 1991) and not to HeI, as previously suggested
by Bakker et~al. (1996a).
We have not found any other HeI, NII or OII lines.
Our analysis shows that the T$_{eff}$ is 8500$\pm$500K.

Bakker et al. (1996b)
found small amplitude light and velocity variations and suggested that 
HD~101584 is a binary with an orbital period of 218 days. The 
radial velocity variations may be due to pulsation, macroturbulence
motions or shock waves in the outer layers of the stellar atmosphere.
Many post-AGB supergiants show small amplitude
light and velocity variations (Hirvnak 1997). These variations
may not be interpreted as due to the presence of a binary companion.
Long term monitoring of the radial velocities is needed in order
to understand the causes for these variations. 

       The spectrum and the brightness of HD~101584 appears 
to remain the same during last two or three decades. There
is no evidence for significant variations in brightness 
similar to those observed in Luminous Blue Variables (LBVs).
The chemical composition and all the available multiwavelength 
observational data collected during the last two decades
by various observers indicates that HD~101584 is most
likely a post-AGB star.

       The presence of several P-Cygni lines with significant 
outflow velocities,  the OH maser and CO emission profiles
(Te Lintel Hekkert et al. 1992, Trams et al. 1990) and 
the IRAS infrared fluxes and colours (Parthasarathy and
Pottasch 1986) indicates the possibility that HD~101584
is a post-AGB star with a bipolar outflow with a dusty disk.
Since HD~101584 shows a strong H$\alpha$ emission line,
high resolution imaging with the Hubble Space Telescope (HST)
may reveal the bipolar nebula and the presence of a dusty disk
similar to that observed in other post-AGB stars like IRAS 17150-3224
(Kwok et~al. 1998)  or IRAS 17441-2411 (Su et~al. 1998).

\setcounter{figure}{0}
\newpage
\vspace{2cm}
\begin{figure*}
\resizebox{17cm}{!}{\includegraphics{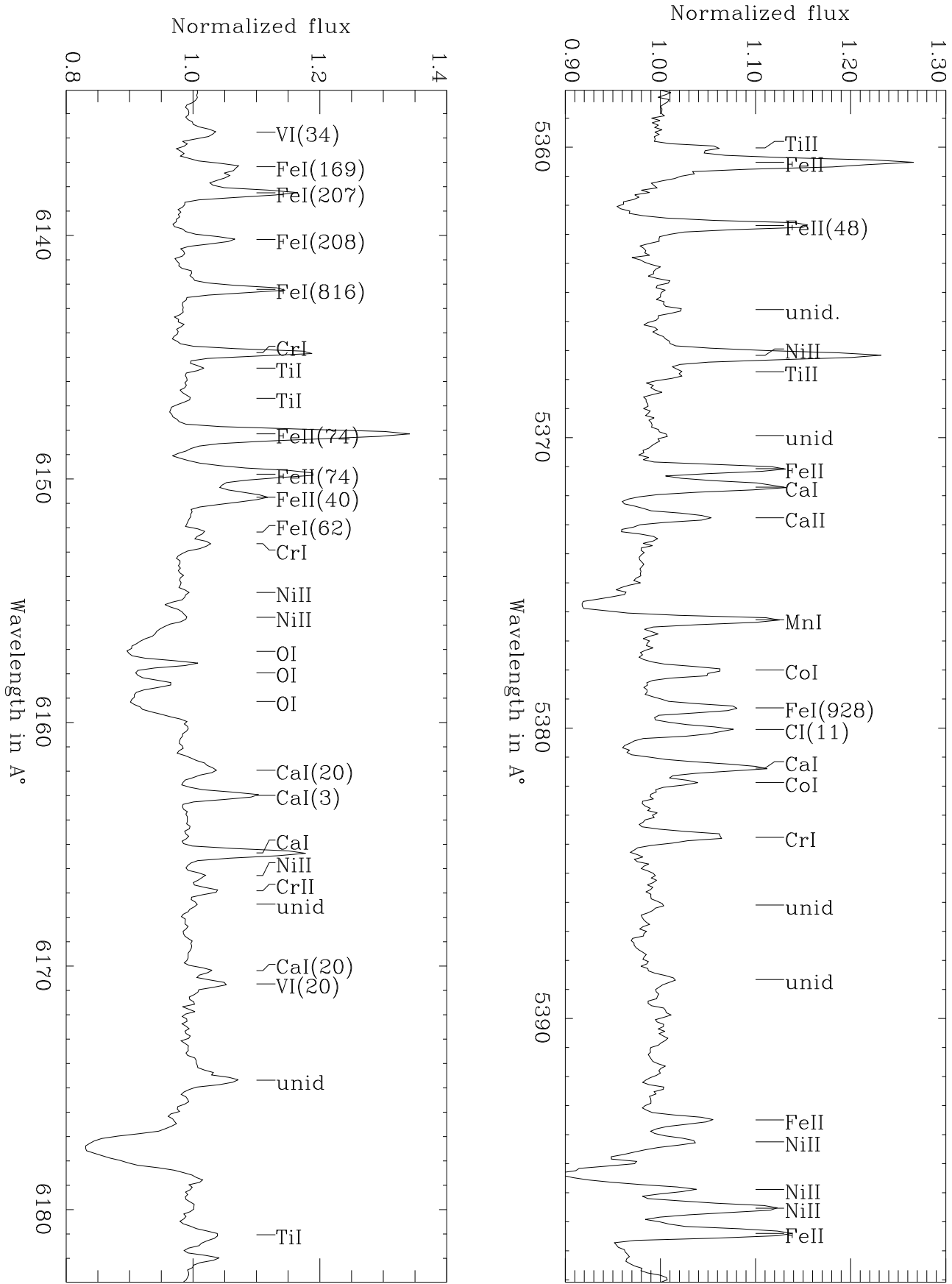}}
\caption{High resolution spectra of HD~101584 obtained with the  ESO CAT-CES}
\label{Fig. 1}
\end{figure*}

\newpage
\setcounter{figure}{0}
\renewcommand{\thefigure}{\arabic{figure}(contd.)}
\begin{figure*}
\resizebox{17cm}{!}{\includegraphics{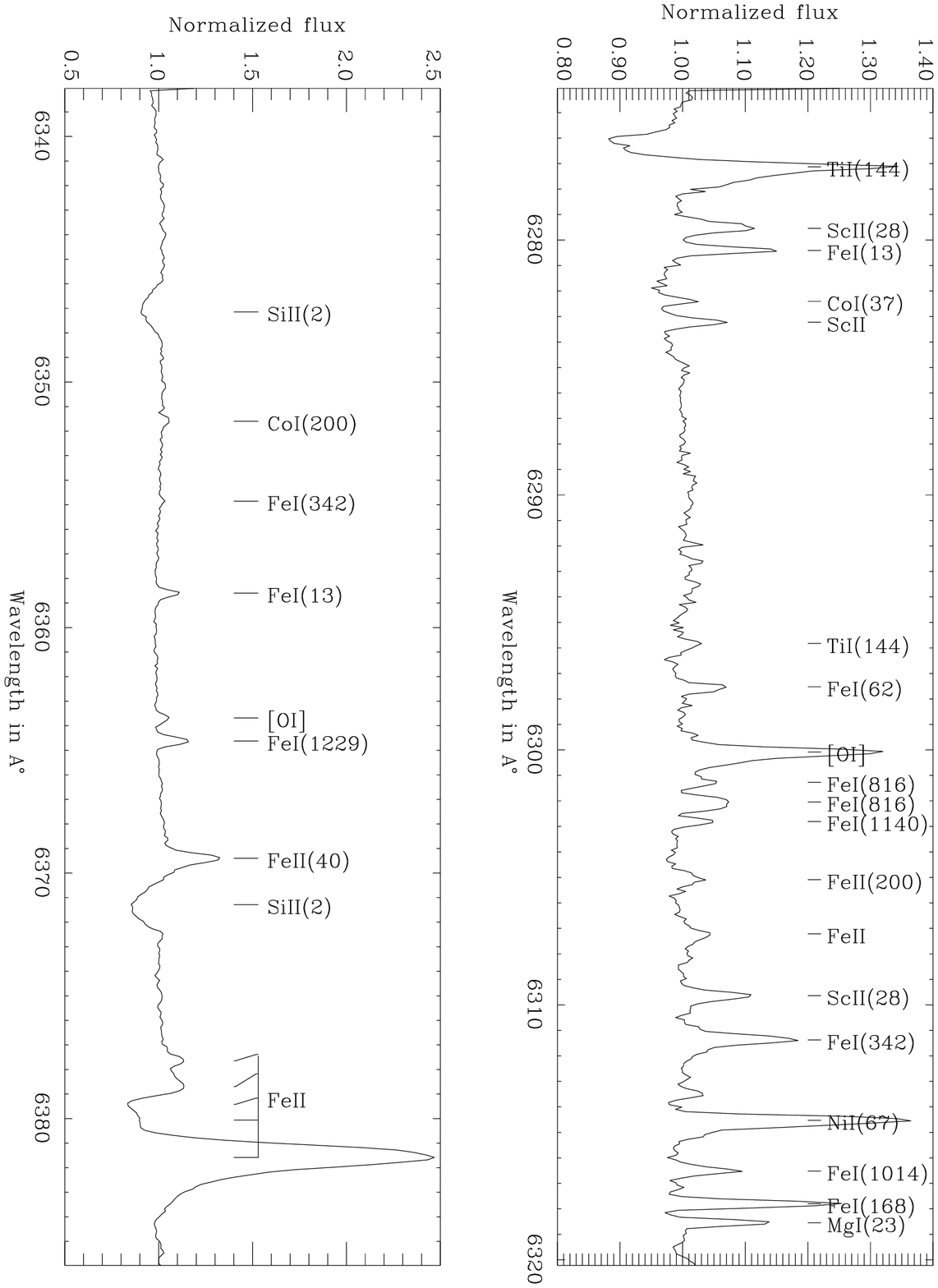}}
\caption{High resolution spectra of HD~101584 obtained with the  ESO CAT-CES}
\label{Fig. 1 (continued)}
\end{figure*}

\newpage
\setcounter{figure}{0}
\vspace{2cm}
\begin{figure*}
\resizebox{17cm}{!}{\includegraphics{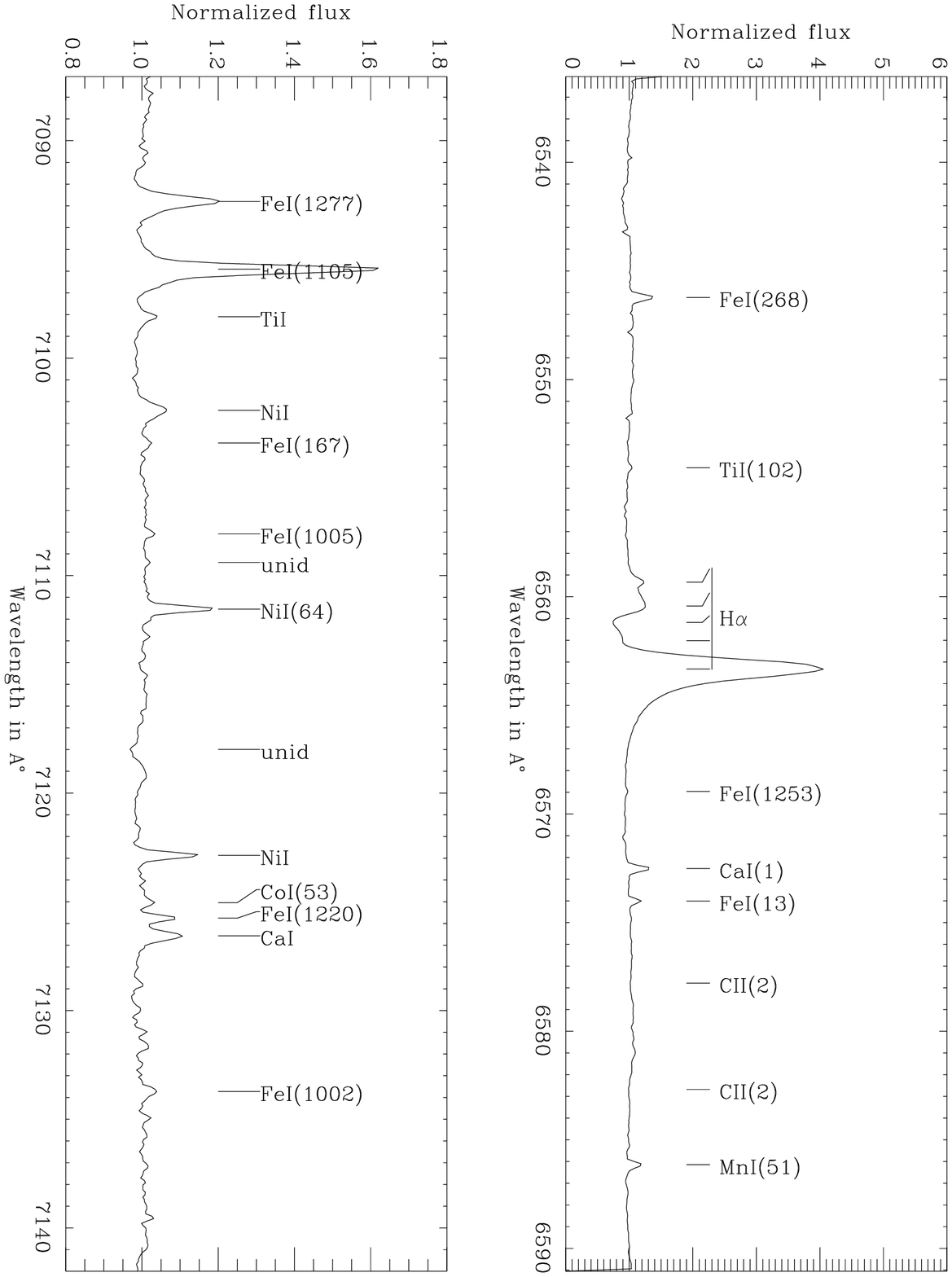}}
\caption{High resolution spectra of HD~101584 obtained with the ESO CAT-CES}
\end{figure*}

\newpage
\setcounter{figure}{0}
\begin{figure*}
\resizebox{17cm}{!}{\includegraphics{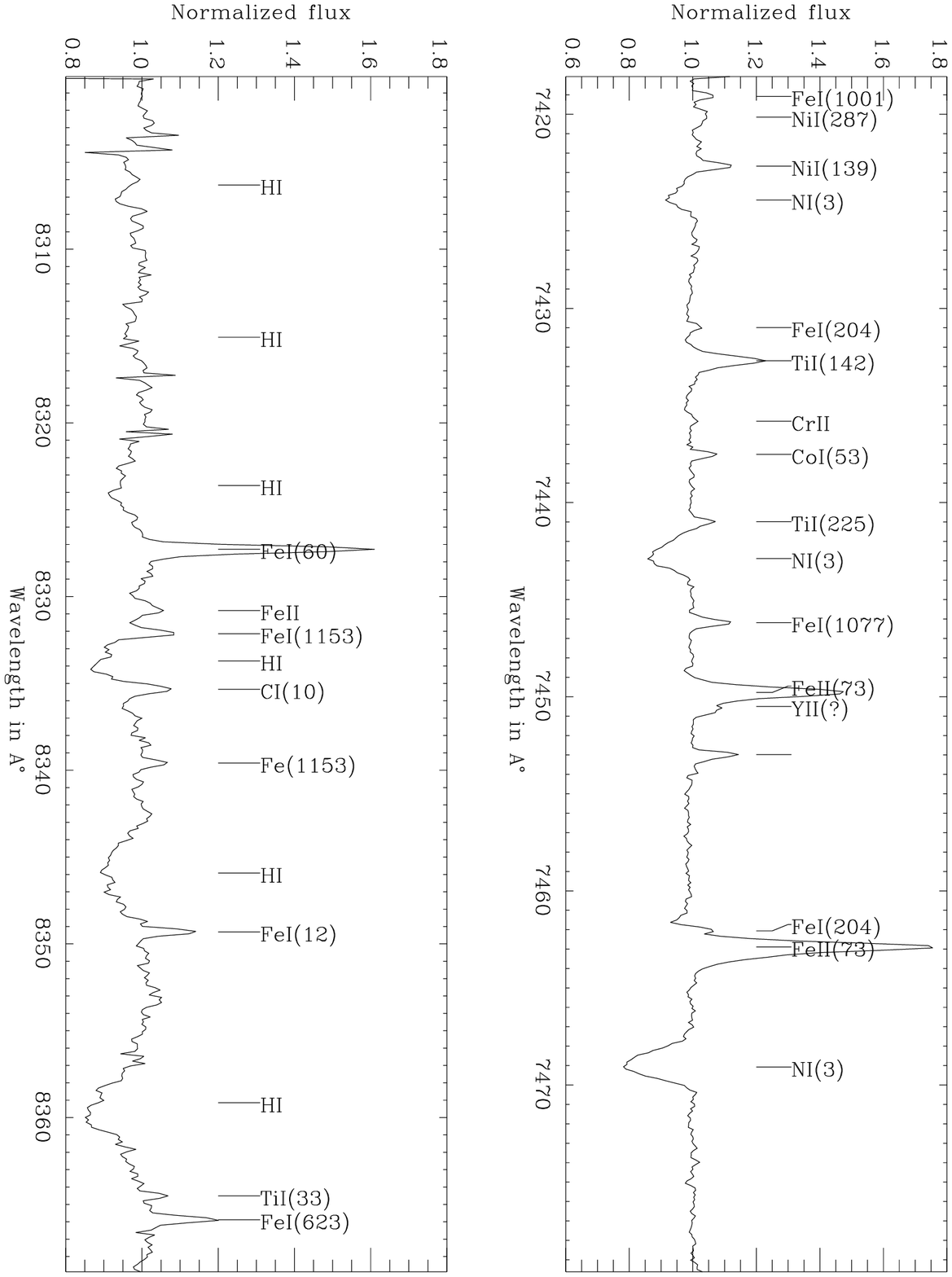}}
\caption{High resolution spectra of HD~101584 obtained with the ESO CAT-CES}
\end{figure*}

\newpage
\setcounter{figure}{0}
\begin{figure*}
\resizebox{8.5cm}{!}{\includegraphics{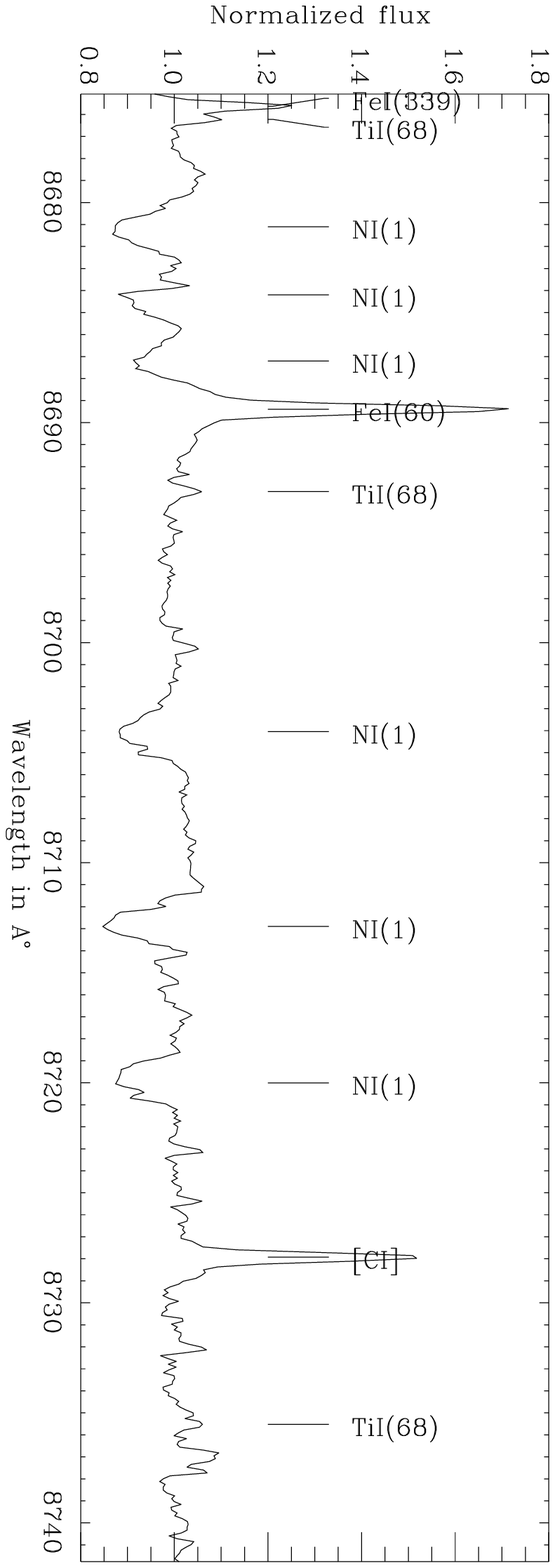}}
\caption{High resolution spectra of HD~101584 obtained with the ESO CAT-CES}
\end{figure*}

\newpage
\setcounter{figure}{1}
\renewcommand{\thefigure}{\arabic{figure}}
\begin{figure*}
\resizebox{17cm}{!}{\includegraphics{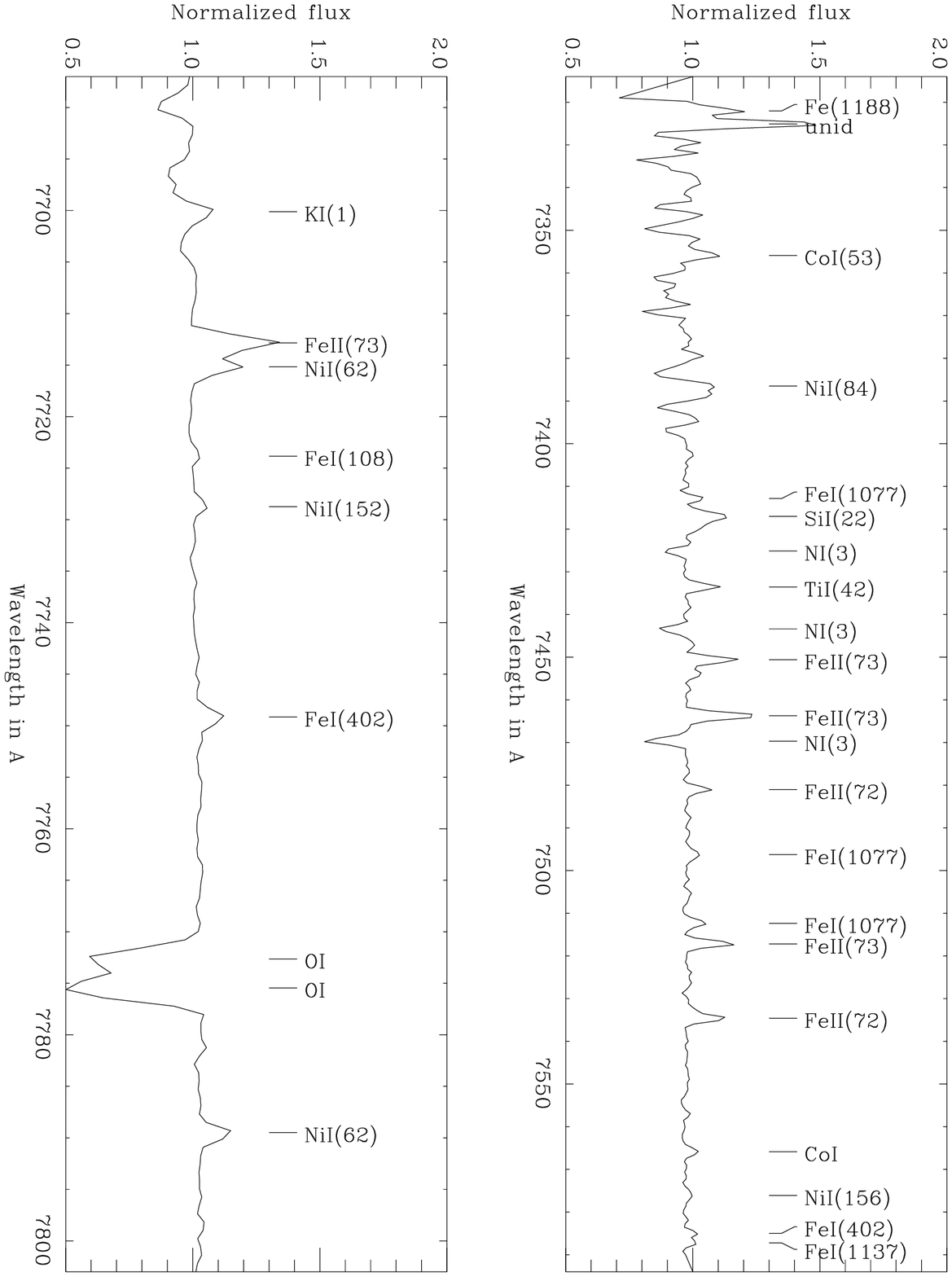}}
\caption{\em The spectrum in the upper panel shows several nitrogen lines
and emission lines of Fe. The lower panel shows the KI 7699\AA\ in
emission and the strong absorption due to  OI triplet at 7777\AA.}
\end{figure*}

\newpage
\begin{figure*}
\vspace{1cm}
\resizebox{17cm}{!}{\includegraphics{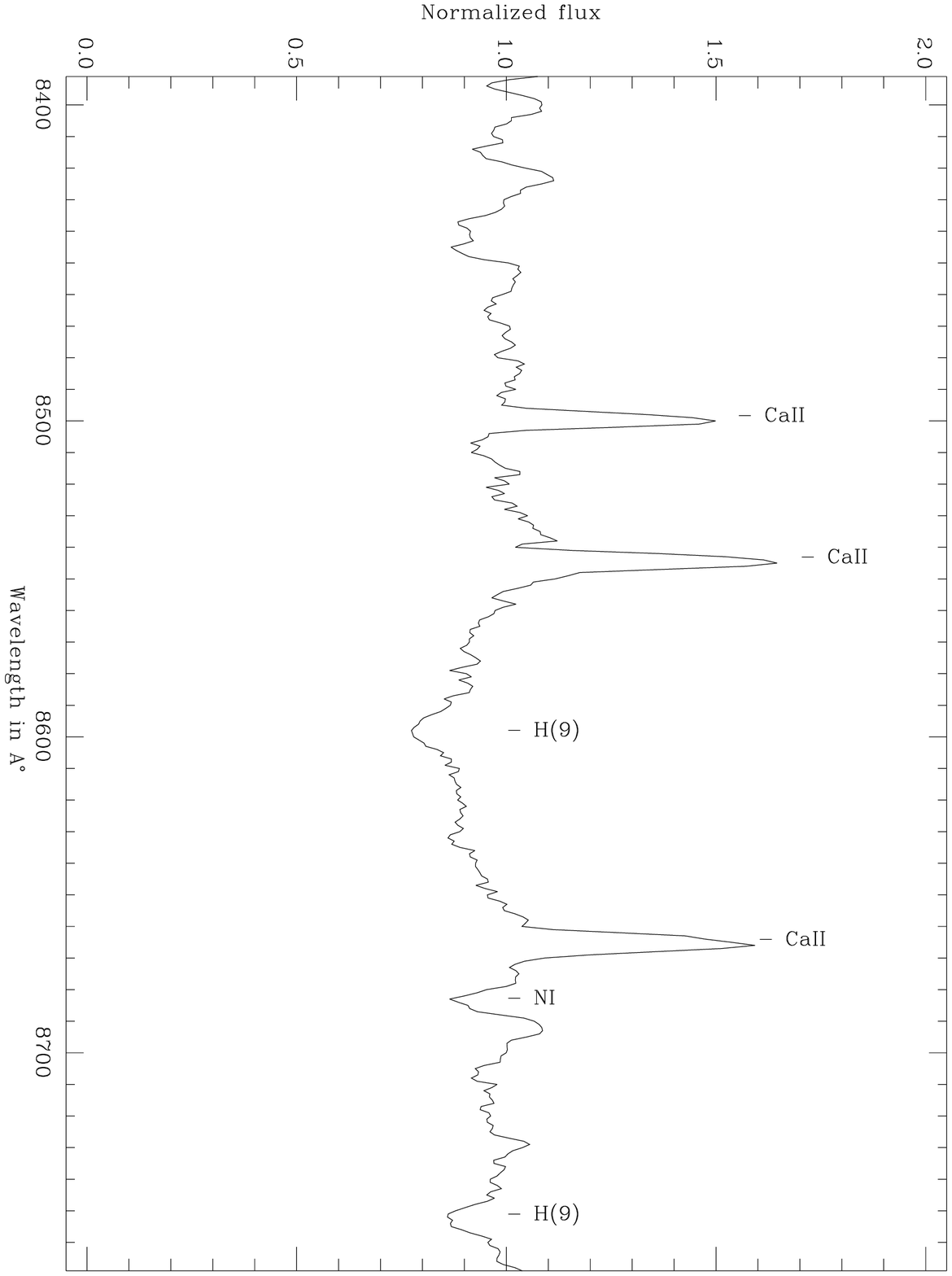}}
\caption{\em CaII IR triplet lines showing P-Cygni emission.This spectrum is
of 2.5\AA\ resolution, obtained from VBO, Kavalur.}
\end{figure*}

\newpage
\begin{figure*}
\vspace{1cm}
\resizebox{17cm}{!}{\includegraphics{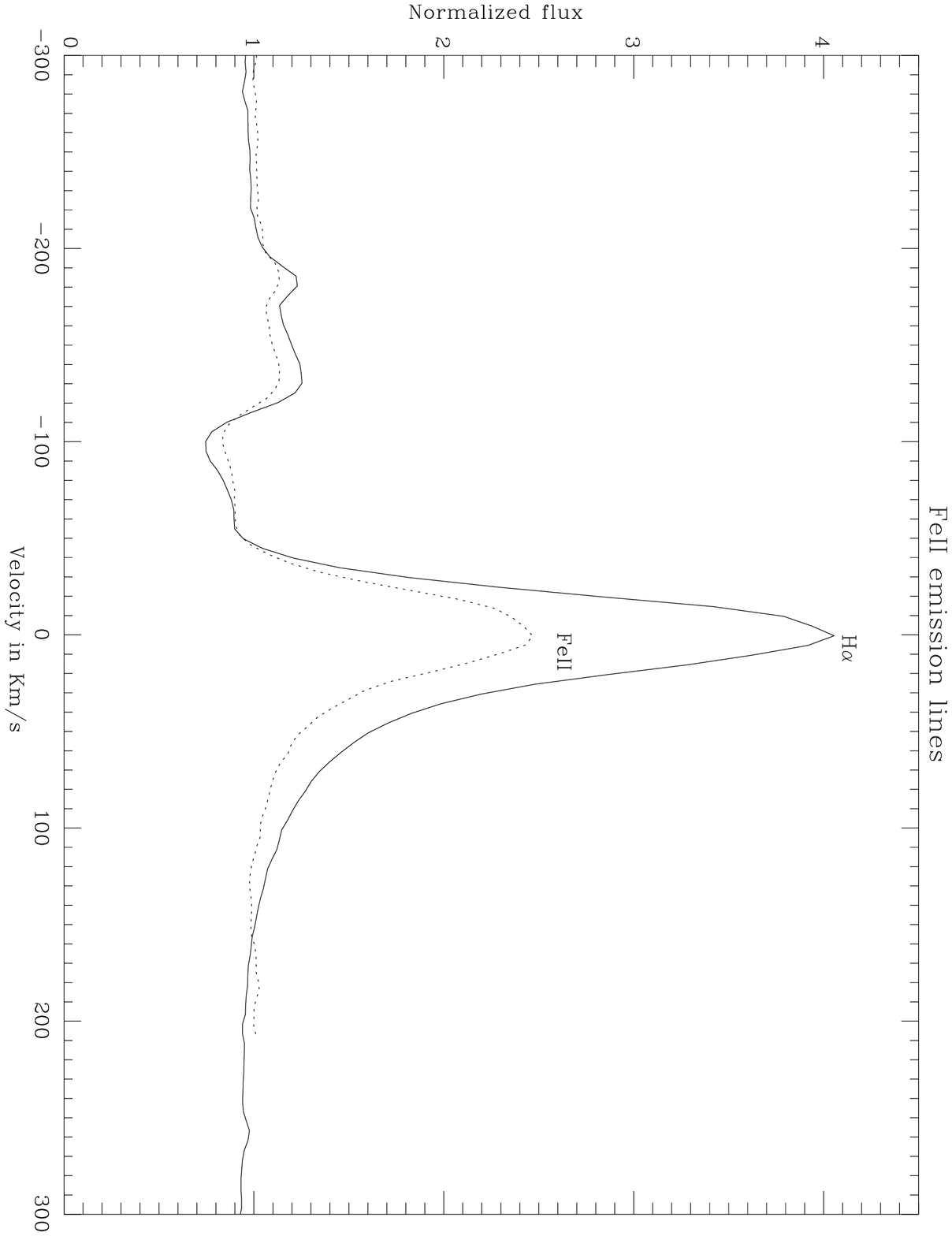}}
\caption{\em P-Cygni profile of H$\alpha$ and FeII(6383\AA) lines showing similar
velocity structures}
\end{figure*}

\newpage
\begin{figure*}
\vspace{1cm}
\resizebox{17cm}{!}{\includegraphics{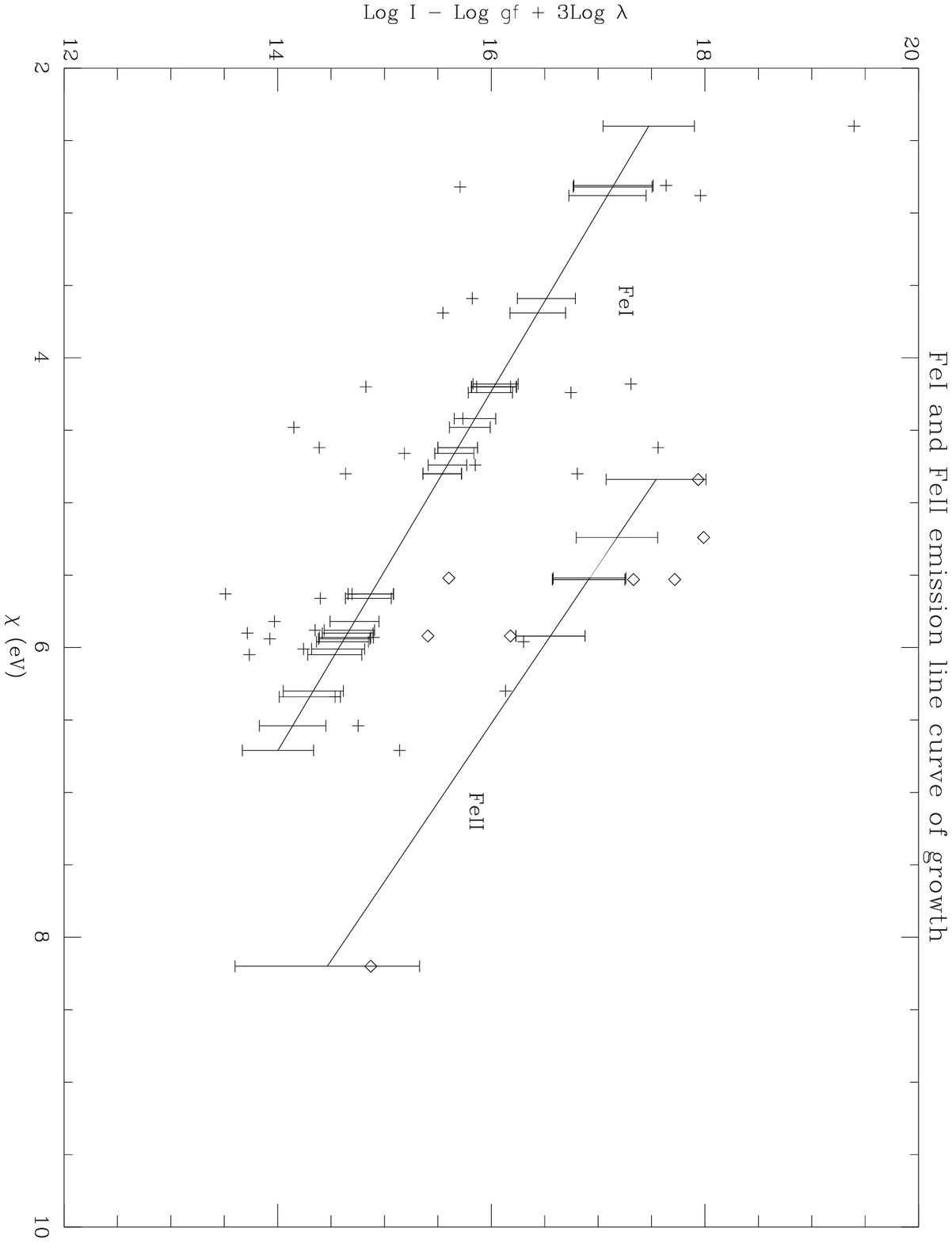}}
\caption{\em Curve of growth analysis of Fe emission lines. + represents the 
FeI lines and $\diamondsuit$ represents FeII lines. The slope gives 
T$_{exi}$=6300$\pm$1000K for FeI lines and T$_{exi}$=5550$\pm$1700K for the
FeII lines. The large dispersion is because the lines are optically thick.
The errors bar show the error in the least square fit. }
\end{figure*}

\newpage
\setcounter{figure}{5}
\renewcommand{\thefigure}{\arabic{figure}(a)}
\begin{figure*}
\vspace{1cm}
\resizebox{17cm}{!}{\includegraphics{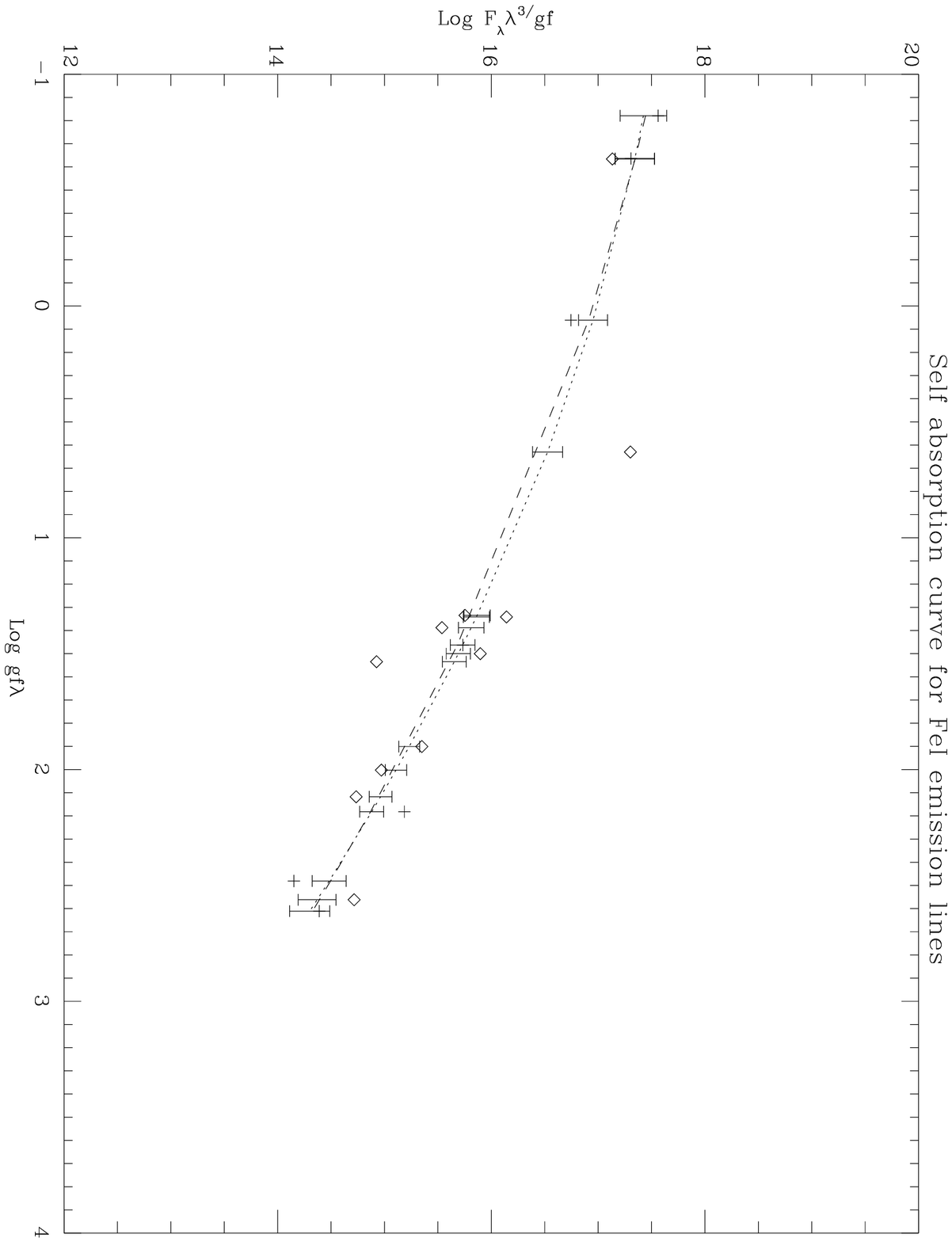}}
\caption{\em The plot shows the shape of the SAC. The + sign indicates
multiplets 167,168,169,204,207,208 of Fe I having similar excitation potential. 
$\diamondsuit$ indicates multiplets 1002,1005,1014,1077,1105,1140,1153,1220,1229,1277
of FeI. The fit was obtained after shifting 
higher multiplets 1002,1005,1014,1077,1105,1140,1153,1220,1229,1277 w.r.to the 
lower multiplets 167,168,169,204,207,208.}
\end{figure*}

\newpage
\setcounter{figure}{5}
\begin{figure*}
\vspace{1cm}
\resizebox{17cm}{!}{\includegraphics{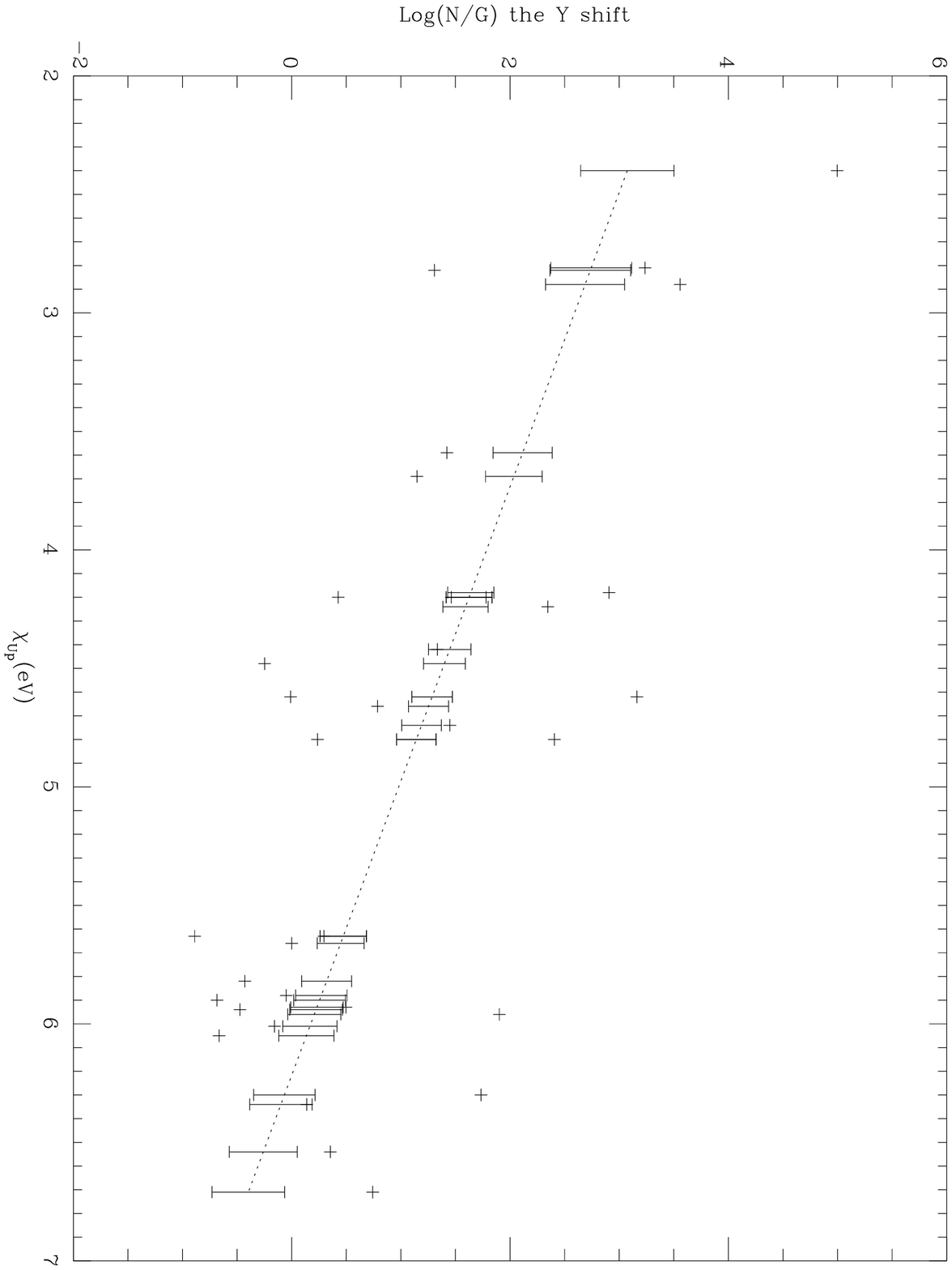}}
\renewcommand{\thefigure}{\arabic{figure} (b)}
\caption{
The fit shows the distribution of upper  level
population of different multiplets of FeI with respect to the 
multiplet 207, versus the upper 
excitation potential.}
\end{figure*}

\newpage
\setcounter{figure}{5}
\renewcommand{\thefigure}{\arabic{figure} (c)}
\begin{figure*}
\vspace{1cm}
\resizebox{17cm}{!}{\includegraphics{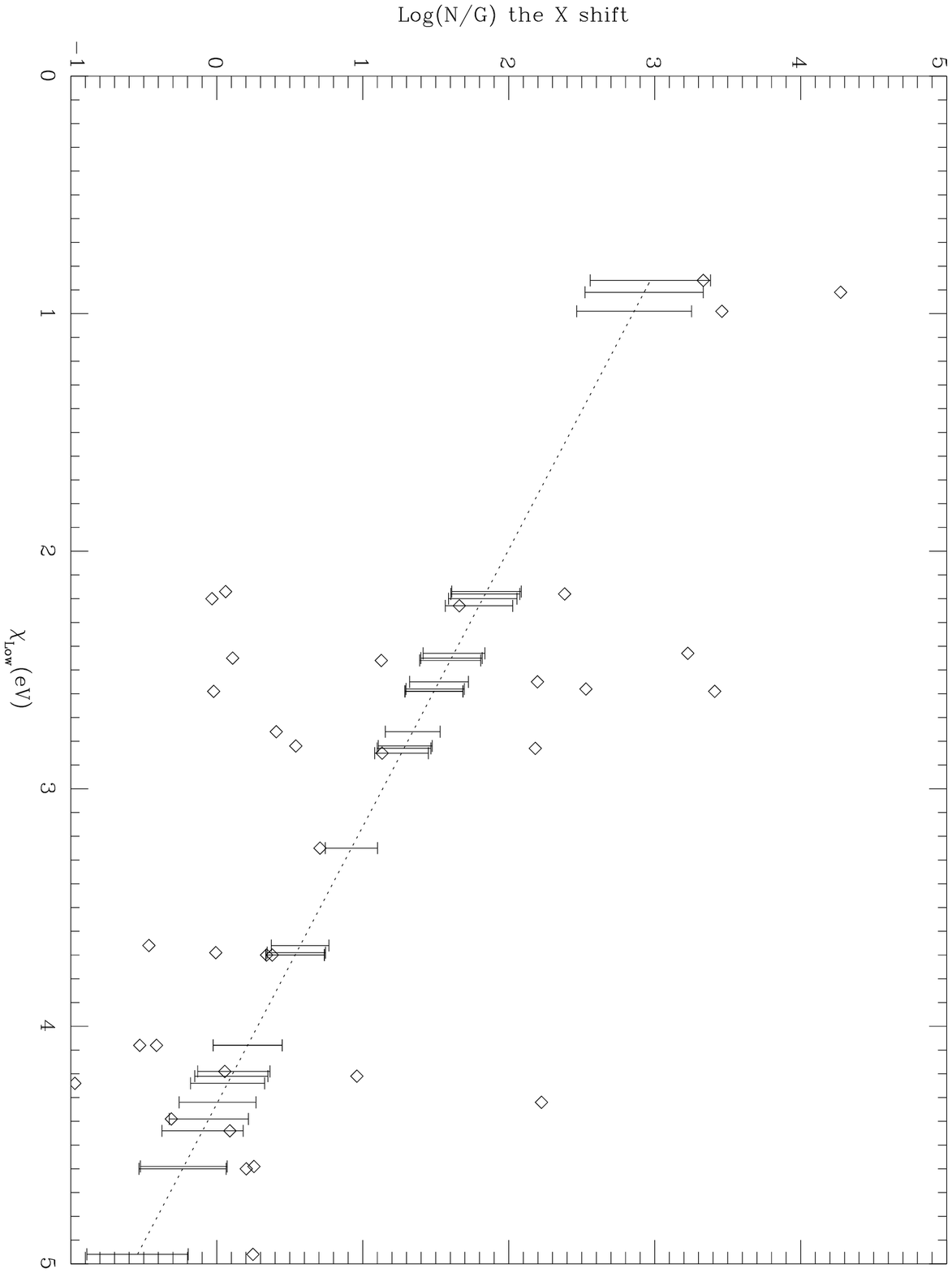}}
\caption{
The fit shows the distribution of  lower level
population of different multiplets of FeI with respect to the 
multiplet 207, versus the  lower
excitation potential.}
\end{figure*}

\newpage
\setcounter{figure}{6}
\renewcommand{\thefigure}{\arabic{figure}}
\begin{figure*}
\caption{\em Plot of electron density N$_{e}$ and electron 
temperature T$_{e}$.  The dotted line is the contour for the 
observed ratio (13.3) of [O]I lines 5577\AA, 6300\AA\ and 6363\AA. 
Each contour in the plot is for a change in the flux ratio of 0.5. }
\resizebox{17cm}{!}{\includegraphics{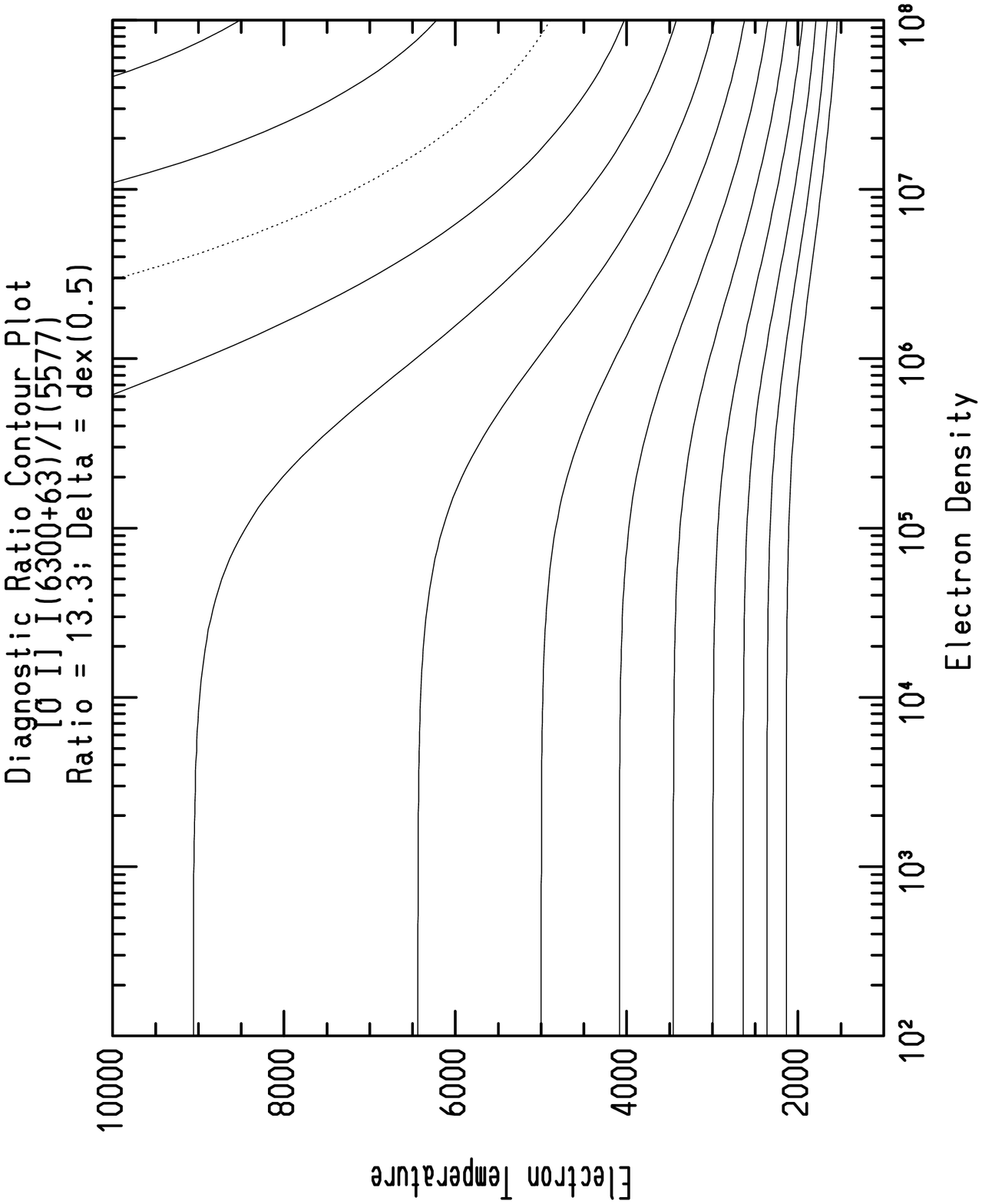}}
\end{figure*}

\newpage
\begin{figure*}
\vspace{1cm}
\resizebox{17cm}{!}{\includegraphics{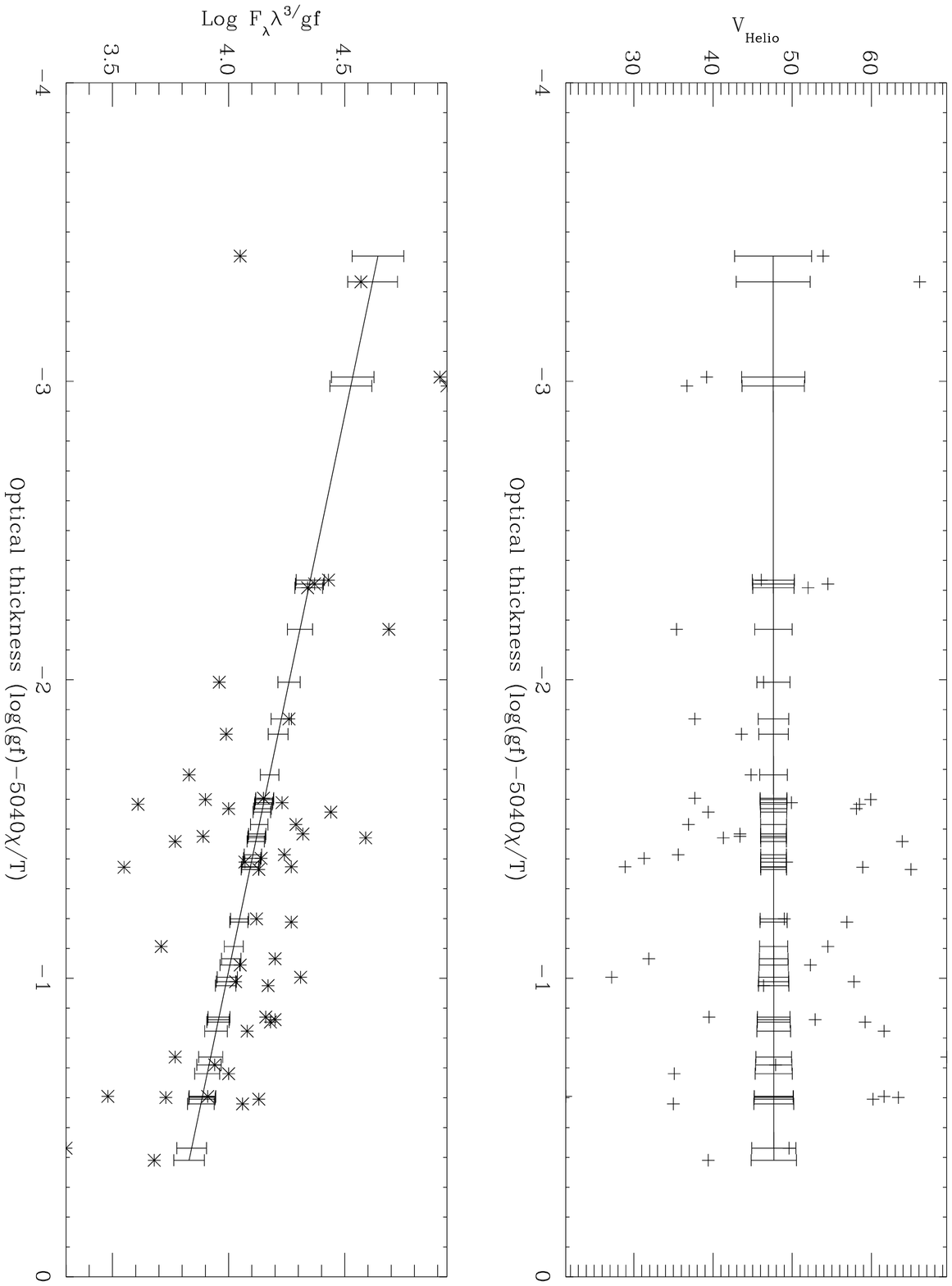}}
\caption{\em The plot in the upper panel does not show any correlation between 
the optical depth and the helio centric radial velocity for the FeI 
absorption lines of HD~101584 in the wavelength region 3600\AA-4500\AA.
The plot in the lower panel shows the normalized strength of FeI absorption
lines in the wavelength region 3600\AA\ to 4500\AA\ versus the optical depth.
It shows clearly that the lines are forming at different optical depths. 
The equivalent widths  are taken from the paper by Rosenzweig et~al.(1997)}
\end{figure*}

\newpage
\begin{figure*}
\vspace{1cm}
\resizebox{17cm}{!}{\includegraphics{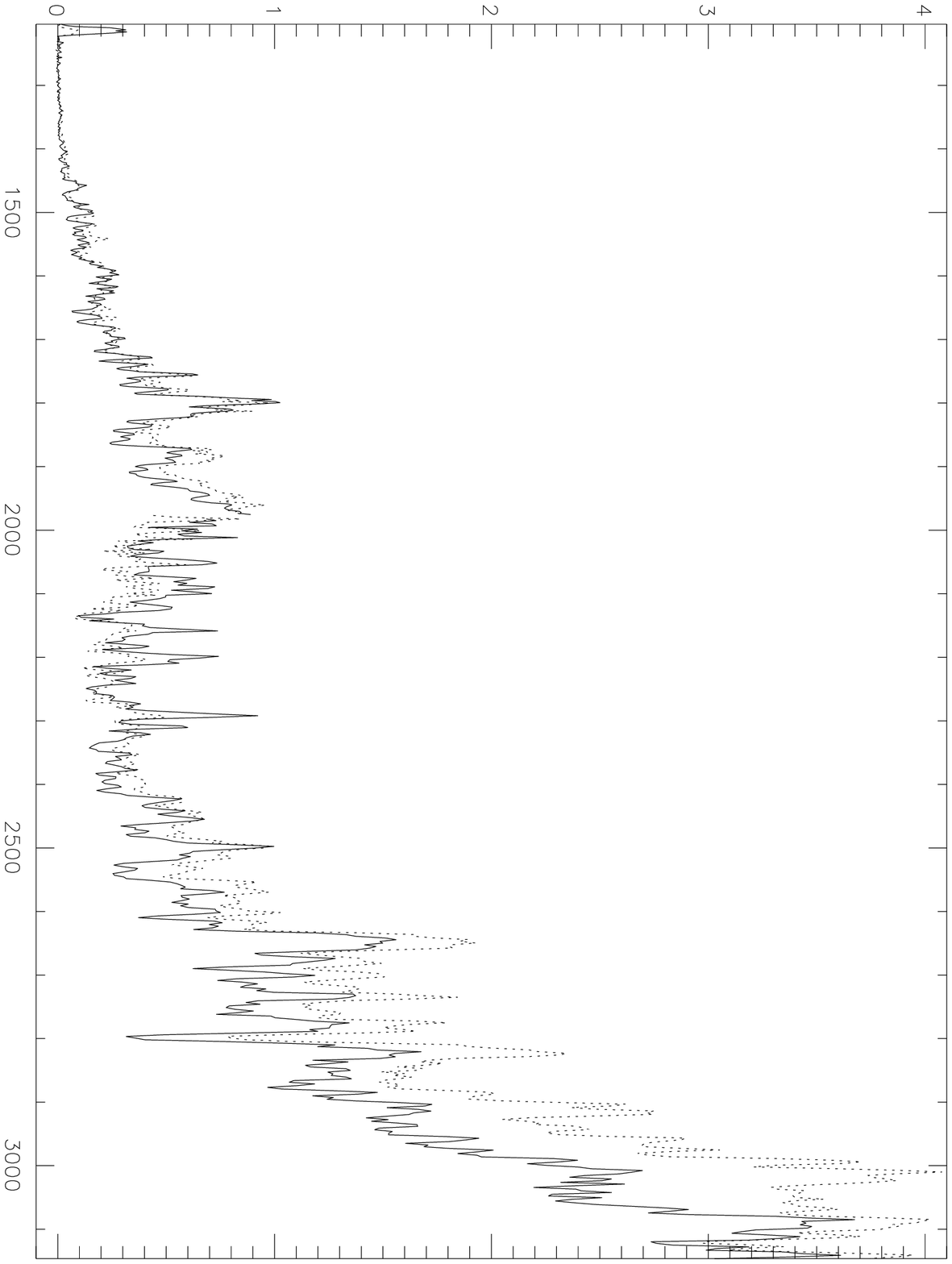}}
\caption{\em IUE low resolution spectrum of HD~101584 
is compared with that the A6Ia star HD~97534.  The dotted
line corresponds to the spectrum of HD~97534 while the solid
line is HD~101584.}

\end{figure*}

\newpage
\begin{figure*}
\vspace{1cm}
\resizebox{17cm}{!}{\includegraphics{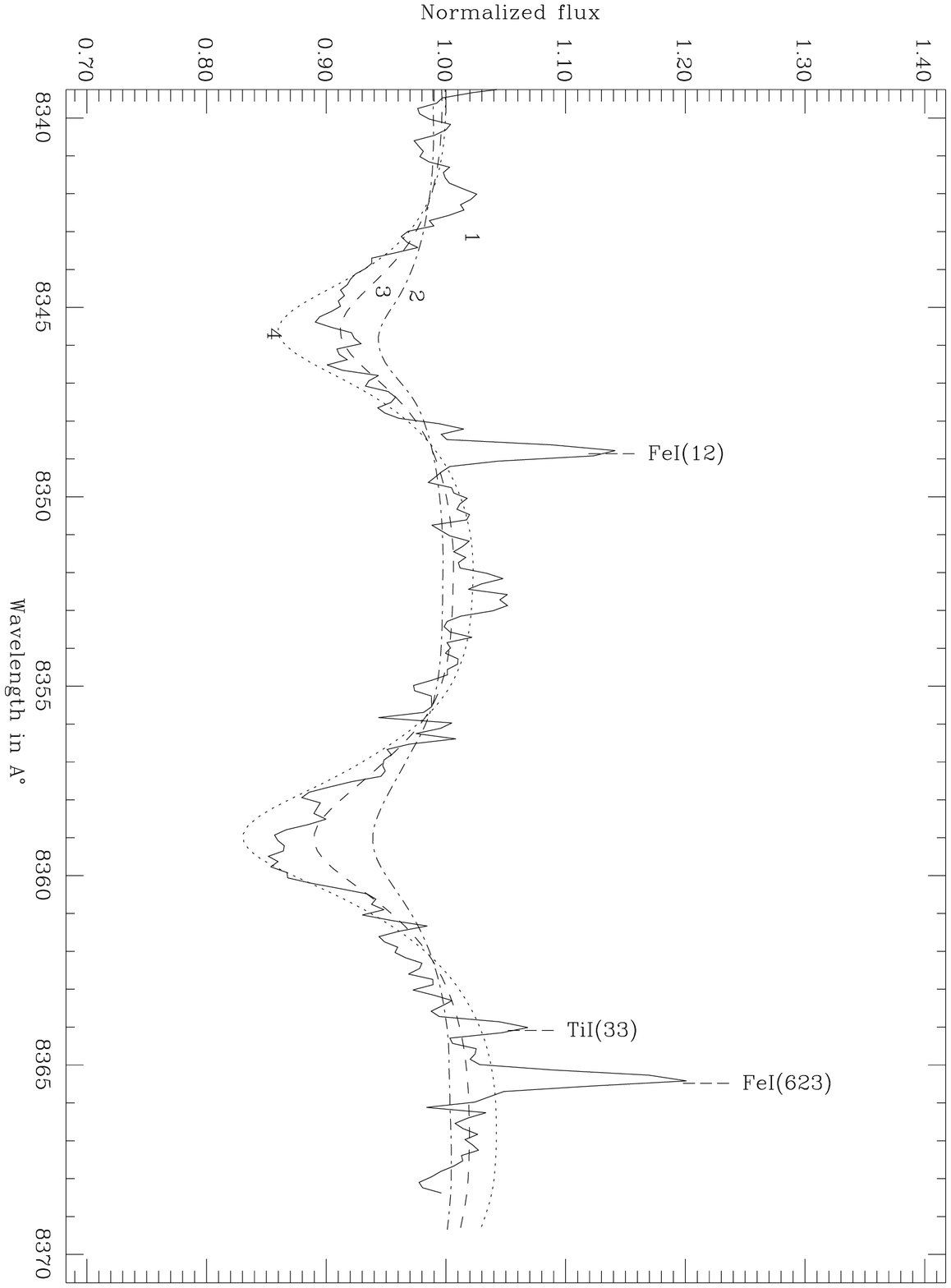}}
\caption{\em Observed and synthetic spectra in the
 Paschen line region. 1-observed, 2-Teff=8000K,log g=1.0,
3-Teff=8000K,log g=2.0, 4-Teff=8500K, log g=1.5. The peaks are FeI emission lines}
\end{figure*}

\newpage
\begin{figure*}
\vspace{1cm}
\resizebox{17cm}{!}{\includegraphics{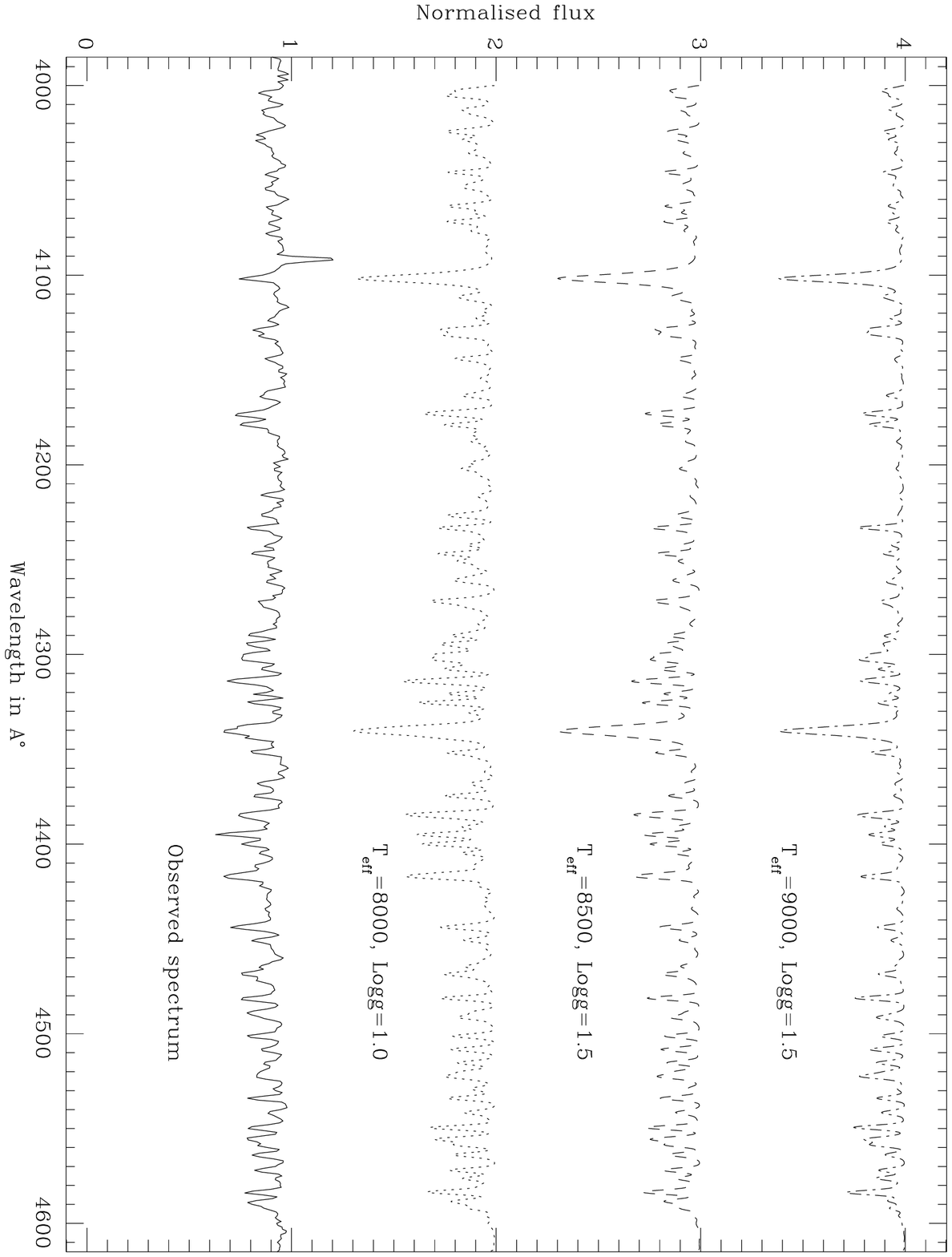}}
\caption{\em Synthesis spectra for different models are compared 
with the observed spectrum. The observed spectrum is of 2.5\AA\ resolution
taken at VBO kavalur. Observed spectrum matches well for T$_{eff}$=8500K, 
log g=1.5, V$_{turb}$=13km~s$^{-1}$
and [Fe/H]=0.0. }
\end{figure*}

\begin{thebibliography}{}
\bibitem{} Bakker E.J., 1994, A\&AS 103, 189
\bibitem{} Bakker E.J.,Lamers H.J.G.L.M., Waters L.B.F.M., Waelkens C.,Trams 
N,R., Van Winckel H., 1996a, A\&A 307, 869
\bibitem{} Bakker E.J.,Lamers H.J.G.L.M., Waters L.B.F.M., Waelkens C., 1996b, 
A\&A 310, 861
\bibitem{} Friedjung M., Muratorio G. 1987, A\&A, 188, 100
\bibitem{} Garc\'\i a-Lario P., Parthasarathy M., de Martino D., Monier R.,
Manchado A., de C\'ordoba S.F., Pottasch S.R., 1997, A\&A, 326, 1103
\bibitem{} Garc\'\i a-Lario P., Manchado A., Parthasarathy M., Pottasch S.R., 1994, A\&A,
285, 179
\bibitem{} Hibbert A., Bi\'emont E., Godefroid M., Vaeck N., 1991, A\&AS, 88, 505
\bibitem{} Hrivnak B., 1997, in Planetary Nebulae, ed. H. J.Habing and 
H.J.G.M.  Lamers, IAU symp. No. 180, 303
\bibitem{} Hoffleit D., Saladya M., Wlasuk P., 1983, Supplement to the Bright star 
Catalogue, Yale University Observatory, USA
\bibitem{} Hubeny I., Stefl S., Harmanec P., 1985, Bull. Astron. Inst.
 Czechosl. 36, 214
\bibitem{}Humphreys R.M., 1976, ApJ 206, 122
\bibitem{} Humphreys R.M., Ney E.P., 1974, ApJ 190, 339
\bibitem{} Kurucz R.L., 1994, Solar Abundance Model Atmospheres, Kurucz CD-ROM No.19,
Smithsonian Astrophysical Observatory
\bibitem{} Kwok S., Su K.Y.L., Hrivnak B.J.,  1998, ApJ, 501, L117
\bibitem{} Lang K.R., 1992, Astrophysical data: Planets and stars, Springer Verlag
\bibitem{} Morrison N.D., Zimba J.R., 1989, BAAS 21, 1022
\bibitem{} Osterbrock D.E., 1989, Astrophysics of Gaseous Nebulae and 
Active Galactic Nuclei, Oxford University Press, p.117
\bibitem{}Parthasarathy M., Garc\'\i a-Lario P., Pottasch S.R., 1992, A\&A,
264, 159
\bibitem{} Parthasarathy M., Pottasch S.R., 1986, A\&A, 154, L16
\bibitem{} Reddy B.E., Parthasarathy M., Gonzalez G., Bakker E.J.,
1997, A\&A 328, 331
\bibitem{} Rosenzweig P., Reinoso E.G., Naranjo O., 1997, JRASC, 91, 255
\bibitem{} Su K.Y.L., Volk K., Kwok S., Hrivnak B.J., 1998, ApJ 508, 744
\bibitem{}Te Lintel Hekkert P., Chapman J.M., Zijlstra A.A., 1992, ApJ 390, L23
\bibitem{}Trams N.R., Van der Veen W.E.C.J., Waelkens C., Waters L.B.F.M. 1990, 
A\&A 233, 153
\bibitem{} Viotti R., 1969, Astrophys. Space Sci. 5, 323
\bibitem{} Wiese W.L., Smith M.W., Glennon B.M., Atomic Transition Probabilities,
 Vol. 1, NSRDS-NBS(U.S.) 4, 1966
\bibitem{} Wiese W.L., Martin G.A., Wavelength and Transition Probabilities for
Atoms and Atomic Ions, NSRDS-NBS(U.S.), 68, 1980
\end{thebibliography}
\end{document}


\setlength{\voffset}{-0.8truein}
\setcounter{page}{23}
\begin{table*}
\caption{ List of emission lines detected in the high resolution spectrum
of HD~101584}
\begin{tabular}{llllllll}  \hline
$\lambda$ obs  & $\lambda$ lab  &Ident. &~~~~~~$\chi$~~~~~~~& log  gf & F$_{\lambda}$$^\ast$ & $\Delta$$\lambda$  & V$_{Helio}$  \\
(\AA) &  (\AA) & &~~~(eV)~~~&  & (m\AA) & (\AA) &  km/s \\
\hline 
5360.022 &  5360.115 &  TiII &2.60-4.92 &-4.040&& -0.093  & 10.60 \\  
5360.517 &  5360.755 &  FeII &10.52-12.83 &-3.082&118.4& -0.237 &  2.51  \\ 
5362.698 &  5362.864 &  FeII(48) &3.20-5.52&-2.739&47.36&  -0.165 &  6.56 \\  
5365.597 &   &  unid  & &&&  &  \\ 
5367.167 &  5367.205 &  NiII  &14.61-16.93&-2.571&79.71&-0.038  & 13.69  \\ 
5367.731 &  5367.912 &  TiII  &1.57-3.88&-4.480&& -0.181 &  5.69  \\ 
5369.928 &  5369.928 &  unid  &&&&   &   \\ 
5371.07  &  5371.275 &  FeII  &10.56-12.87&-0.705&45.52& -0.20  & 4.36  \\ 
5371.704 &  5371.934 &  CaI   &5.4-7.71&-3.163&&-0.220  & 2.97  \\ 
5372.76  &  5372.807 &  CaII &9.03-11.35&-1.337&21.64 & -0.04  & 13.17  \\ 
5376.27  &  5376.477 &  MnI &2.93-5.23&-4.091&45.13&  -0.20 &  4.26  \\ 
5378.    &  5378.247 &  CoI &4.03-6.34 &-0.425&32.41& -0.24 &  2.03  \\ 
5379.305 &  5379.581 &  FeI(928) &3.70-6.01&-1.521&33.83&  -0.275 &  0.43 \\  
5380.048 &  5380.322 &  CI(11) &7.70-10.00&-1.840&32.67&  -0.273  & 0.54  \\ 
5381.379 &  5381.599 &  CaI &5.46-7.77&-4.960&49.97&  -0.220  & 3.54  \\ 
5381.874 &  5382.046 &  CoI &4.51-6.82&-3.234&15.82&  -0.171  & 6.24  \\ 
5383.763 &  5383.927 &  CrI &4.54-6.85 &-2.606&22.12& -0.163  & 6.70  \\ 
5386.093 &  5386.093 &  unid &&&&     &  \\ 
5388.659 &  5388.659 &  unid & &&&   &   \\ 
5393.491 &  5393.762 &  FeII &11.19-13.49&-2.154&17.12&  -0.270  & 0.74  \\ 
5394.247 &  5394.496 &  NiII &12.64-14.94&-1.029&14.61&  -0.249  & 1.97  \\ 
5395.886 &  5396.051 &  NiII &14.63-16.93&-2.851&10.14 & -0.164  & 6.67 \\  
5396.532 &  5396.701 &  NiII  &14.63-16.93&-2.561&54.78& -0.168 &  6.42  \\ 
5397.4   &  5397.640 &  FeII  &10.73-13.03&-2.883&51.29& -0.240 &  2.46  \\ 
6133.723 &  7132.985 &  FeI(1002) &4.08-5.82&-0.851&36.31&  0.738 &  46.96   \\
6135.76  &  6135.370 &  VI(34) &1.05-3.08&-0.750&23.13&  0.389  & 35.22 \\  
6137.179 &  6136.624 &  FeI(169) &2.45-4.48&-1.307&30.29&  0.555 &  43.31 \\  
6138.25  &  6137.702 &  FeI(207) &2.59-4.62&-1.177&70.63&  0.547 &  42.95  \\ 
6140.167 &  6139.65  & FeI(208)  &2.59-4.62&-4.609&38.79& 0.517  & 41.43  \\ 
6142.214 &  6141.727 &  FeI(816) &3.70-5.63&-1.540&61.82&  0.486 &  39.95  \\ 
6154.659 &  6154.094 &  NiII &14.88-16.91&-2.268&5.54&  0.564 &  43.71  \\ 
6144.813 &  6144.294 &  CrI  &4.10-6.13 &-3.624&80.21&0.519  & 41.51  \\ 
6145.448 &  6144.936 &  TiI  &1.89-3.91&-2.520&16.08& 0.512 &  41.17  \\ 
6146.688 &  6146.27  & TiI(153)  &3.18-5.20&-2.541&15.79& 0.417  & 36.57 \\  
6148.146 &  6147.742 &  FeII(74)  &3.89-5.92&-2.721&20.85& 0.403  & 35.87 \\  
6149.804 &  6149.249 &  FeII(74)  &3.89-5.92&-2.724&122.7& 0.555  & 43.25  \\ 
6150.747 &  6150.10  & FeII(40)  &3.22-5.24&-4.754&73.71& 0.646  & 47.73  \\ 
6152.178 &  6151.624 &  FeI(62) &2.18-4.20&-3.582&17.06&  0.554 &  43.20  \\ 
6152.655 &  6152.098 &  CrI  &4.10-6.12&-1.986&33.16& 0.556  & 43.31  \\ 
6155.679 &  6155.102 &  NiII &13.14-15.16&-0.682&7.19&  0.577 &  44.30  \\ 
6161.957 &  6161.295 &  CaI(20) &2.53-4.54&-1.293&27.97&  0.662  & 48.41 \\  
6162.99  &  6162.180 &  CaI(3)  &1.90-3.92&-0.167&49.42& 0.810  & 55.61 \\  
6165.354 &  6164.716 &  CaI  &6.05-8.07&-2.261&73.94& 0.638  & 47.22  \\ 
6166.282 &  6165.893 &  NiII &9.75-11.77&-2.297&9.67&  0.389 &  35.10  \\ 
6166.907 &  6166.187 &  CrII &13.06-15.07&-3.490&18.24&  0.720 &  51.21  \\ 
6167.456 &  6167.456 &  unid &&&10.3&   &    \\ 
6170.189 &  6169.559 &  CaI(20) &2.53-4.54&-0.527&13.17&  0.629 &  46.80 \\  
6170.733 &  6170.340 &  VI(20) &4.80-6.81&-0.654&23.53&  0.216 &  26.71 \\  
6174.685 &  6174.685 &  unid  &&&39.17&   &   \\ 
6181.032 &  6180.625 &  TiI &3.18-5.19&-3.632&43.54&  0.407 &  35.93 \\  
6277.13  &  6277.470 &  TiI(144) &1.73-3.71&-3.794&112.2&  -0.205  & 6.20 \\  
6279.541 &  6279.740 &  ScII(28) &1.5-3.48&-1.265&41.49&  -0.287 &  2.29 \\  
6280.406 &  6280.622 &  FeI(13) &0.86-2.83&-3.720&44.44&  -0.216 &  5.67 \\  
6282.404 &  6282.638 &  CoI(37)  &1.74-3.72&-2.021&13.03& -0.234  & 4.81\\   
\hline
\end{tabular}
\end{table*}

\newpage
\setcounter{table}{0}
\begin{table*}
\renewcommand{\thetable}{\arabic{table} (contd.)}
\caption{List of emission lines detected in the high resolution spectrum
 of HD~101584 }

\begin{tabular}{llllllll} 
\hline 
$\lambda$ obs  & $\lambda$ lab  &ID &~~~~~~$\chi$~~~~~~~& log gf & F$_{\lambda}$$^\ast$  & $\Delta$$\lambda$  & V$_{Helio}$  \\
(\AA) &  (\AA) & &~~~~(eV)~~~&  & (m\AA) & (\AA) &  km/s \\
 \hline 

6283.225 &  6283.353 &  ScII &7.45-9.43&-0.441&40.34&  -0.127 &  9.90 \\  
6295.826 &  6295.949 &  TiI(144)& & & & & \\
6297.524 &  6297.799 &  FeI(62) &2.23-4.20&-2.871&39.23&  -0.274 &  2.91 \\  
6300.083 &  6300.311 &  [OI]  &&&163& -0.228  & 5.15 \\  
6301.27  &  6301.508 &  FeI(816) &3.66-5.63&-0.745&23.32&  -0.237  & 4.68  \\ 
6302.042 &  6302.499 &  FeI(816) &3.69-5.66&-1.203&62.88&  -0.457  & -5.75\\   
6302.807 &  6303.461 &  FeI(1140)&4.32-6.30&-3.434&20.01&   -0.653 &  -15.11 \\  
6305.09  &  6305.314 &  FeII(200)&6.23-8.20&-2.039&27.17&   -0.224 &  5.34 \\  
6307.214 &  6307.529 &  FeII  &2.83-4.80&-5.685&57.4& -0.314  & 1.02 \\  
6309.636 &  6309.886 &  ScII(28) &1.5-3.47&-1.630&55.71&  -0.250 &  4.12 \\  
6311.379 &  6311.504 &  FeI(342) &2.83-4.80&-3.392&103.2&  -0.125 &  10.06 \\  
6313.52  & &&&&&&\\
6314.529 &  6314.668 &  NiI(67) &4.16-6.13&-0.921&163.5&  -0.139 &  9.39 \\  
6316.52  &  6315.814 &  FeI(1014)&4.08-6.05&-0.683&44.72&   0.706 &  49.54\\   
6317.802 &  6318.027 &  FeI(168)  &2.46-4.42&-2.338&98.8& -0.181  & 7.40 \\  
6318.545 &  6318.717 &  MgI(23)   &5.12-7.08&-1.730&60.1&-0.171 &  7.84  \\ 
6351.598 &  6351.448 &  CoI(200) &&&&&\\
6354.852 &  6355.027 &  FeI(342) &2.85-4.80&-2.346&7.6&  -0.174 &  7.62  \\ 
6358.596 &  6358.687 &  FeI(13)  &0.86-2.81&-4.546&47.9& -0.0908 &  11.59  \\ 
6363.68  &  6363.79  & [OI]  &&&28.9 &-0.109  & 10.69 \\  
6364.625 &  6364.706 &  FeI(1229) &4.59-6.54&-1.469&74.78&  -0.081 &  12.05 \\  
6369.4   &  6369.464 &  FeII(40) &2.89-4.84&-4.253&186.8&  -0.063  & 12.86 \\  
6380.058 &  6383.715 & FeII &5.56-7.50&-2.271&1512&  -3.656  & -155.97 \\  
6381.582 &  6381.416 &  TiI(196) &&&&&\\
6539.783 &  6539.72  &  FeI(405) &&&&&\\
6546.217 &  6546.245 &  FeI(268) &2.76-4.66&-1.634&127.2 & -0.0283  & 14.97 \\  
6554.056 &  6554.226 &  TiI(102)  &1.44-3.34&-1.201&32.48& -0.169  & 8.49 \\  
6561.179 &  6561.179 &  H$\alpha$  &&&3681&  &  \\  
6563.326 &  6563.403 &  CoI(80)&&&&&\\
6568.959 &  6569.231 &  FeI(1253) &&&&&\\
6572.512 &  6572.779 &  CaI(1) &0.00-1.89&-4.104&124.5&  -0.266 &  4.10 \\  
6574.009 &  6574.238 &  FeI(13)&0.99-2.88&-4.688&65.55&   -0.229 &  5.82 \\  
6574.831 &  6575.022 &  FeI(207) &&&&&\\
6580.992 &  6581.220 &  FeI(34) &&&&&\\
6586.144 &  6586.343 &  MnI(51)&&&&&\\ 
6587.438 &  6587.75  &  CI(22)&&&&&\\ 
7092.793 &  7091.942 &  FeI(1277)&4.96-6.71&-1.509&120.4 &  0.886 &  53.22 \\  
7095.905 &  7095.425  & FeI(1105) &4.21-5.96&-2.221&337.4&  0.479 &  36.21 \\  
7098.1   &  7097.655   &TiI  &3.30-5.30&-3.423&16.88 &0.445 &  34.74 \\  
7102.397 &  7101.932  & NiI  &4.54-6.29&-1.941&53.25& 0.464  & 35.55 \\  
7103.904 &   7103.15 & FeI(167)&2.43-4.18 &-4.488&18.39&   &   \\ 
7108.084 &  7107.468  & FeI(1005)  &4.19-5.94&-1.317&11.32& 0.449  & 34.87 \\  
7109.397 &  7109.397  & unid &&&8.62&    &  \\  
7111.541 &  7110.905  & NiI(64)  &1.93-3.68&-3.042&70.85& 0.648  & 43.29 \\  
7117.991 &  7117.991  & unid &&&23.54&    &   \\ 
7122.867  & 7122.191  & NiI(126)  &3.54-5.29&-0.169&55.08& 0.676  & 44.40  \\ 
7125.038  & 7124.47  &CoI(53)  &&&14.34&    &   \\ 
7125.752 &  7125.283  & FeI(1220)  &4.60-6.34&-1.465&32.71& 0.468  & 35.65  \\ 
7126.566 &  7126.19   &CaI  &6.02-7.77&-0.647&60.47 &0.375   &31.74  \\ 
7419.082 &  7418.668 &  FeI(1001)&4.12-5.79 &-1.17&107.8&  0.414 &  32.62  \\ 
7420.15  &  7419.31  & NiI(287)  &5.50-7.17&-0.695&& 0.839  & 49.84  \\ 
7422.671 &  7422.30  & NiI(139) &3.69-5.28&+0.06&57.36&  0.371 &  30.88 \\  
\hline
\end{tabular}
\end{table*}

\newpage
\setcounter{table}{0}
\begin{table*}
\renewcommand{\thetable}{\arabic{table} (contd.)}
\caption{ List of emission lines detected in the high resolution spectrum
of HD~101584 }

\begin{tabular}{llllllll}  \hline
$\lambda$ obs  & $\lambda$ lab  &ID &~~~~~~$\chi$~~~~~~~& log gf & F$_{\lambda}$$^\ast$  & $\Delta$$\lambda$  & V$_{Helio}$  \\
(\AA) &  (\AA) & &~~~~(eV)~~~&  & (m\AA) & (\AA) &  km/s \\
\hline

7430.991 &  7430.5  & FeI(204) &2.58-4.24&-3.81&21.02&  0.491 &  35.71 \\  
7432.698 &  7431.97 &  TiI(142) &1.74-3.41&-2.283&15.53&  0.728&   45.27 \\  
7435.814 &  7435.3  & unid  &&&47.41&   &   \\ 
7437.521 &  7437.16 &  CoI(53) &1.95-3.61&-3.64&45.96&  0.360 &  30.43 \\  
7440.984 &  7440.6  & TiI(225) &2.25-3.90&-1.19&38.7&  0.383 &  31.35  \\ 
7446.188  & 7445.70 &  FeI(1077) &4.24-5.90&-0.31&61.72&  0.487  & 35.53 \\  
7449.784  & 7449.34 &  FeII(73) &3.87-5.53&-3.60&316.5&  0.444  & 33.77 \\  
7450.505  & 7450.33 &  YII(?)  &&&& 0.174  & 22.92 \\  
7452.982  & 7452.50& FeII(14F) &&&54.88&   &   \\ 
7462.061  & 7461.527 &  FeI(204) &2.55-4.20&-3.48&&  0.534 &  37.36 \\  
7462.889  & 7462.38 &  FeII(73) &3.87-5.53&-2.98&540.5 & 0.509 &  36.35 \\  
8327.273  & 8327.061&   FeI(60) &2.20-3.69&-1.298&307.8&  0.212 &  23.79 \\  
8330.807  & 8330.587 &  unid && &55.59&  &   \\ 
8332.142  & 8331.926 &  FeI(1153) &4.39-5.88&-1.020&36.98&  0.215 &  23.89 \\  
8335.342  & 8335.150 &  CI(10) &7.70-9.19&-0.420&69.24&  0.191  & 23.01  \\ 
8339.584  & 8339.398 &  Fe(1153) &4.44-5.93&-1.421&51.58&  0.185 &  22.80  \\ 
8349.31   & 8349.02  & FeI(12)  &0.91-2.40&-5.605&106.4& 0.290  & 26.54  \\ 
8364.508  & 8364.243 &  TiI(33) &0.83-2.32&-1.652&28.96&  0.264 &  25.61  \\ 
8365.885  & 8365.642 &  FeI(623)&3.25-4.74&-2.040&110.0&   0.243 &  24.84  \\ 
8675.626  & 8674.751 &  FeI(339)&2.82-2.42 &-1.89&100.7&  0.875 &  46.10  \\ 
8676.22   & 8675.38  & TiI(68)  &1.06-2.50&-1.357&43.2& 0.839   &44.88  \\ 
8689.396  & 8688.632 &  FeI(60) &2.17-3.59&-1.41&393.9&  0.764 &  42.24\\   
8693.139  & 8692.34  & TiI(68) &1.04-2.46&-1.92&27.48&  0.798  & 43.41  \\ 
8727.923 &  8727.4   &[CI]  &&&302.5& 0.522  & 33.80  \\ 
8735.521  & 8734.70  & TiI(68) &1.05-2.48&-2.087&&  0.821  & 44.05 \\  
\hline
\end{tabular}
\end{table*}

\newpage
\renewcommand{\thetable}{\arabic{table}}
\begin{table*}
\caption{Chemical composition of HD101584. }
\begin{tabular}{l|l|l|l}
\hline\hline
Element&\multicolumn{2}{c|}{[element/H]} &Lines\\ \cline{2-3}
&T$_{eff}$=8500K& T$_{eff}$=9000K&\\
&log g=1.5&log g=1.5&\\
\hline
C & 1.0$\pm$0.1 & 0.3$\pm$0.1 &C II 6578,6582\\
&&&\\
N & 0.5$\pm$0.1 & 0.6$\pm$0.1 & N I 7423,7442,8703,\\
&&&8711,8718,8728\\
&&&\\
O &0.1&0.18&O I 6158\\
&&&\\
Mg$^{1}$&0.2&0.02&Mg II 4481.13\\
Ti$^{1}$&0.0$\pm$0.4&0.4$\pm$0.4&Ti II 15 lines\\
Fe$^{1}$&-0.1$\pm$0.1&0.5$\pm$0.1&Fe I 3 lines\\
&0.2$\pm$0.4&0.3$\pm$0.4&Fe II 6lines\\
\hline
\hline
\end{tabular}
\end{table*}